# Unraveling Joint Evolution of Bars, Star Formation, and Active Galactic Nuclei of Disk Galaxies

Woong-Bae G. Zee,[1,2] Sanjaya Paudel,[1,2] Jun-Sung Moon,[3,4] and Suk-Jin Yoon[1,2]

[1]*Department of Astronomy, Yonsei University, Seoul, 03722, Republic of Korea*
[2]*Center for Galaxy Evolution Research, Yonsei University, Seoul, 03722, Republic of Korea*
[3]*Astronomy Program, Department of Physics and Astronomy, Seoul National University, Seoul, 08826, Republic of Korea*
[4]*Research Institute of Basic Sciences, Seoul National University, Seoul, 08826, Republic of Korea*

## ABSTRACT

We aim to unravel the interplay between bars, star formation (SF), and active galactic nuclei (AGNs) in barred galaxies. To this end, we utilize the SDSS DR12 to select a sample of nearby ($0.02 < z < 0.06$) disk galaxies that are suitable for bar examination ($M_r < -20.12$ and inclination $\lesssim 53°$). We identify 3662 barred galaxies and measure the length and axis ratio of each bar. We invent new bar parameters that mitigate the stellar and bulge mass biases and show, for the first time, that the evolution of non-AGN and AGN-hosting barred galaxies should be tracked using different bar parameters; the bar length for non-AGN galaxies and the bar axis ratio for AGN-hosting galaxies. Our analysis confirms that barred galaxies have a higher specific SF rate than unbarred control galaxies. Moreover, we find a positive correlation of bar length with both the SF enhancement and the centrally star-forming galaxy fraction, indicating the interconnectivity of bars and SF through the bar-driven gas inflow. We also find that while the AGN fraction of barred galaxies is the same as that of the unbarred control sample, galaxies hosting more massive black holes (BHs) have rounder (i.e., higher axis ratio) bars, implying that the bar is not a cause of AGN activity; rather, AGNs appear to regulate bars. Our findings corroborate theoretical predictions that bars in non-AGN galaxies grow in length, and bars in AGN-hosting galaxies become rounder as BHs grow and eventually get destroyed.

## 1. INTRODUCTION

The bar is one of the common ($\sim$30%) features of disk galaxies (e.g., Eskridge et al. 2000; Knapen et al. 2000; Reese et al. 2007; Menéndez-Delmestre et al. 2007; Marinova & Jogee 2007; Barazza et al. 2008; Aguerri et al. 2009; Masters et al. 2011; Lee et al. 2012a; Simmons et al. 2014; Erwin 2018; Kim et al. 2020; Géron et al. 2021). Due to gas inflow driven by the non-axisymmetric potential (Sellwood & Wilkinson 1993; Piner et al. 1995; Sakamoto et al. 1999; Regan et al. 1999; Maciejewski et al. 2002; Regan & Teuben 2004; Kim et al. 2012; Sellwood 2013; Zana et al. 2019), bars are thought to be connected with galactic evolution, through such processes as star formation (SF) (Martinet & Friedli 1997; Allard et al. 2006; Vera et al. 2016; Robichaud et al. 2017; Kim et al. 2017; Fraser-McKelvie et al. 2020a; Wang et al. 2020; Moon et al. 2022; Lu et al. 2022) and/or active galactic nucleus (AGN) activity (Shlosman et al. 1989; Combes 2001; Alonso et al. 2013; Alonso et al. 2014; Galloway et al. 2015; Kim & Choi 2020). However, the exact mechanisms are still controversial. Over decades, both observations and simulations have yielded diverse results.

Whether bars enhance or suppress SF is not clear (e.g., Pompea & Rieke 1990; Martinet & Friedli 1997; Chapelon et al. 1999; Knapen et al. 2006; Cheung et al. 2013; Willett et al. 2015; Kim et al. 2017; Wang et al. 2020). Conventionally, it is thought that barred galax-

Corresponding author: Suk-Jin Yoon
sjyoon0691@yonsei.ac.kr

ies have higher star formation rates (SFRs) than unbarred galaxies, particularly at the center (Athanassoula 1994; Sakamoto et al. 1999; Sheth et al. 2005; Ellison et al. 2011; Coelho & Gadotti 2011; Wang et al. 2012; Zhou et al. 2015; Chown et al. 2019; Lin et al. 2020). Star formation in barred galaxies is typically enhanced along the length of the bar (e.g., NGC 1022 by Garcia-Barreto et al. 1991; NGC 1073 and 3059 by Phillips 1996; NGC 1530 by Zurita & Pérez 2008; NGC 2903 by Leon et al. 2008) and within the circumnuclear ring region (e.g., NGC 6782 by Windhorst et al. 2002; NGC 3081 by Byrd et al. 2006; NGC 1097 by Prieto et al. 2019). Coelho & Gadotti (2011) found an excess of young stellar populations in barred galaxies' bulges, indicating bar-driven central SF enhancement. Using integral field spectroscopy, Chown et al. (2019) showed that barred galaxies have higher central molecular gas concentration, which leads to higher SFRs. Lin et al. (2020) analyzed two-dimensional maps of EW(Hα), EW(Hδ$_A$), and $D_n(4000)$ and found that barred galaxies exhibit higher concentration of SF than unbarred galaxies. On the other hand, it has been shown that on a global scale, barred galaxies have less SF activity than unbarred galaxies (Masters et al. 2012; Spinoso et al. 2017; Khoperskov et al. 2018; George et al. 2019a; Fraser-McKelvie et al. 2020b; Géron et al. 2021; George & Subramanian 2021). Fraser-McKelvie et al. (2020b) found that barred galaxies are more common in redder and passive galaxies with lower gas fractions. George & Subramanian (2021) showed that galaxies that host larger bars tend to have lower SFRs.

Similarly, the interplay between bars and the AGNs is complex (e.g., Moles et al. 1995; McLeod & Rieke 1995; Ho et al. 1997; Hunt & Malkan 1999; Shlosman et al. 2000; Berentzen et al. 2007; Zhang et al. 2009; Oh et al. 2012; Cisternas et al. 2013; Cheung et al. 2015; Robichaud et al. 2017; Martel et al. 2018; Alonso et al. 2018; Kim & Choi 2020; Łokas 2022). According to previous simulations, the bar-driven inflow has been an efficient mechanism for fuelling black holes (BHs) at the galactic center (Combes 2001; Emsellem et al. 2015; Fanali et al. 2015). Some observations showed an increase in the AGN fraction and strength for barred galaxies against their unbarred counterparts, supporting the mechanism of bar-induced AGNs (Alonso et al. 2013; Cisternas et al. 2015; Audibert et al. 2019). However, other studies claimed that there is no direct link between the presence of bars and AGN activity (Hunt & Malkan 1999; Gadotti 2008; Zhang et al. 2009; Lee et al. 2012a; Cisternas et al. 2013; Galloway et al. 2015; Cisternas et al. 2015; Goulding et al. 2017; Silva-Lima et al. 2022). Cisternas et al. (2013) used the *Spitzer* Survey of Stellar Structure in Galaxies (S$^4$G) images to determine bar strength and compared it to X-ray luminosity and the Eddington ratio of AGNs from *Chandra* observations. They found no detectable correlation between the bar and AGN strength. Based on bar classification from Galaxy Zoo 2 (GZ 2), Goulding et al. (2017) showed that large-scale bars do not play a significant role in the accretion rate of AGNs. However, Silva-Lima et al. (2022) found that the accretion rate of AGNs is higher in barred galaxies when they adopted different correlations between BH mass and central stellar velocity dispersion for barred and unbarred galaxies.

To explain these disparate results, recent efforts focused on the time-dependent bar evolution since z∼1 (e.g., Okamoto et al. 2015; Spinoso et al. 2017; Zana et al. 2018; Debattista et al. 2019; Zana et al. 2019; Zhou et al. 2020; Kim et al. 2021; Cavanagh et al. 2022; Reddish et al. 2022; Lee et al. 2022; Bi et al. 2022). The redistribution of angular momentum through bars can enhance SF at the early stage of bar growth (Athanassoula 2005; Athanassoula et al. 2013; Athanassoula 2013; Jang & Kim 2022), but later this process leads to SF quenching due to a rapid gas consumption (Jogee et al. 2005; Sheth et al. 2005; Hunt et al. 2008; Carles et al. 2016; Newnham et al. 2020). Also, Combes (2000) and Bournaud & Combes (2002) simulated that, after the bar-driven sufficient gas infall feeds AGNs, bars become rounder and are self-destroyed on ∼1 Gyr time-scale. More recently, using Illustris-1 and TNG100 simulations, Zhou et al. (2020) showed that the efficiency of SF in barred galaxies declines during the evolution of the bar structure, as the effect of AGN feedback increases. Therefore, the presence of bar structures alone does not adequately reflect bar-driven secular evolution. It needs to be investigated how SF and AGN activities are correlated with the growth and destruction of bars.

Bars grow in length as they evolve (Hernquist & Weinberg 1992; Debattista & Sellwood 1998; Athanassoula 2003; Algorry et al. 2017; Zhao et al. 2020; Łokas 2021), thus, the length of bar structures was usually used as an indicator of how much a bar evolves. The relative bar length, which is defined as the bar's physical length normalized by the size of the galaxy, is a particularly useful measure. It should be noted that the relative size of the bar can depend on factors such as the galaxy's stellar mass, the bulge-to-total mass fraction (B/T), and the morphological type along the Hubble sequence (see the results of Hoyle et al. 2011; Oh et al. 2012; George & Subramanian 2021; Kim et al. 2021). Erwin (2018) and Erwin (2019) analyzed barred galaxies from the S$^4$G and found a two-fold correlation between bar length and galaxy stellar mass: bar length is constant for less mas-sive galaxies ($M_*/M_\odot \leq 10^{10.1}$), while there is a strong increasing trend for massive galaxies ($M_*/M_\odot > 10^{10.1}$). The author also suggested that the dependence of bar size on the morphological type is a side effect of that of stellar mass. Whether the bar size depends on stellar and bulge masses is crucial for studying bar-driven secular evolution because SF and AGN activities are closely correlated with stellar mass.

In this paper, we investigate the interplay of bars with SF and BH for non-AGN and AGN barred galaxies, respectively. Our main questions in this study are: (a) What is the best parameter of the bar strength without the contamination of an undesired selection bias? and (b) How does bar-driven secular evolution interplay with SF and BH activities? To address them, we measure the bar properties such as the length and axis ratio of ~3700 face-on barred galaxies in the local universe (0.02 < z < 0.06) from the SDSS Data Release 12 (DR12). We confirm that bar properties depend on stellar and bulge masses. We then, for the first time, define new parameters of bar strength to disentangle the effect of host galaxy properties.

The paper is organized as follows. In Section 2, we introduce our sample and explain how we estimate the bar properties using our automated scheme. In Section 3, we show the bar properties as functions of stellar and bulge masses, and derive new parameters of the bar strength. In Section 4, we investigate how SF and BH activities depend on the bar properties and how they interplay with each other. In Section 5, we summarize our result in the context of a new concept of the chronological bar-driven secular evolution.

## 2. DATA AND METHODOLOGY

### 2.1. Observational Data

All the galaxies used in this study are selected from the SDSS DR12 database (Alam et al. 2015; Beck et al. 2016) by using the SDSS CasJob. We choose ~76,000 sources classified as galaxy and retrieve their images and spectra. We select sample galaxies in the redshift range 0.02 < z < 0.06 and with $r$-band Petrosian absolute magnitude brighter than –20.12. The redshift and magnitude range corresponds to the completeness limit ($m_r \sim 17$) of the SDSS DR12 spectroscopic catalog at z = 0.06. We select a sample of ~10,000 nearly face-on disk galaxies classified as "spiral galaxy other" in the GZ 2 (Willett et al. 2013), according to the following criteria: (a) galaxies that are flagged as spirals, requiring at least 80% of the volunteer votes for the spiral category after the de-biasing procedure, and (b) galaxies that are classified as "not edge-on" spirals, with more than half of the volunteers voting for the spiral category. Then, we refine our sample by cross-referencing it with the SDSS DR7 (Abazajian et al. 2009) catalog and retain only those galaxies with an apparent axis ratio of $IsoA_r/IsoB_r \geq 0.6$. This cut-off corresponds to an inclination $i \lesssim 53°$, ensuring that our sample predominantly comprises disk galaxies with a nearly face-on orientation, thus facilitating the detection of bars.

The emission line fluxes and stellar kinematics are taken from Sarzi et al. (2006) and Thomas et al. (2013), which are measured from the SDSS spectra using Gas AND Absorption Line Fitting (GANDALF) and penalized PiXel Fitting (pPXF) code (Cappellari & Emsellem 2004). The catalog lists emission line fluxes such as Hα, Hβ, [OIII], and [NII]. To investigate the effect of bar-driven secular evolution on SF and AGNs separately, we divide our sample into non-AGN and AGN host galaxies using the distribution on the BPT diagram (Baldwin et al. 1981; Kauffmann et al. 2003). We restrict our AGN sample galaxies having the Hα, Hβ, [OIII], and [NII] lines with the signal-to-noise ratio (S/N) > 3. Galaxies above the theoretical lower limit for AGN host galaxies expected by Kauffmann et al. (2003) are classified as AGN host galaxies. Galaxies that are below the same limit or do not appear in the BPT diagram due to their feeble emission lines are defined as non-AGN galaxies.

We match our selected sample with the catalog by Chang et al. (2015) to obtain the global specific star formation rates (sSFRs). Chang et al. (2015) combined the SDSS and the Wide-Field Infrared Survey Explorer (WISE) photometry for ~1 million galaxies and constructed a comprehensive catalog of $SFR_{Global}$ and $sSFR_{Global}$ using the MAGPHYS (da Cunha et al. 2008; da Cunha et al. 2015) spectral energy distribution (SED) modeling code. Also, the impact of bar-driven evolution is usually expected to happen in the central region within 1-3 kpc. To examine the central SF enhancement, we use the fiber SFR and sSFR from the MPA/JHU catalog (Brinchmann et al. 2004). This catalog provides $SFR_{Fiber}$ and $sSFR_{Fiber}$ estimated by the reddening-corrected Hα luminosity. Considering the redshift range in this study, the SDSS 3″ fiber exactly covers the physical size of 1-3 kpc.

The size and stellar mass of galaxies are taken from the catalogs by Simard et al. (2011) and Mendel et al. (2014). These catalogs provide extensive photometric data of approximately 660,000 galaxies in the SDSS. In Simard et al. (2011), the bulge and disk components were decomposed using all $ugriz$ wavebands and the GIM2D bulge + disk fitting code. They estimated the half-light radius for each component and the B/T in the $g$- and $r$-bands. Following the result, Mendel et al. (2014) measured the stellar mass of the bulge, disk, and total components separately. They showed that the estimation of the stellar mass is more accurate for face-on galaxies due to lower statistical uncertainty. We only select reliable galaxies within five times the standard deviation based on the correlation between total masses and the sum of the bulge and disk masses.

## 2.2. Bar Classification and Measurements

We identify and measure bar structures automatically. There are several methods that have been used to detect bars in galaxies, including visual inspection of galaxy images (Kormendy 1979; de Vaucouleurs et al. 1991; Martin 1995; Hoyle et al. 2011), photometric decomposition of the surface brightness distribution (Ann & Lee 1987; Aguerri et al. 2000; Aguerri et al. 2003; Reese et al. 2007; Méndez-Abreu et al. 2017), and tracing the variation of the position angle (PA) and ellipticity ($\epsilon$) of galaxy isophotes (Wozniak et al. 1995; Laine et al. 2002; Jogee et al. 2004; Marinova & Jogee 2007; Barazza et al. 2008; Aguerri et al. 2009; Barazza et al. 2009; Li et al. 2011; Oh et al. 2012; Díaz-García et al. 2016; Consolandi 2016; Erwin 2018; Yu et al. 2022; Tawfeek et al. 2022; Guo et al. 2022). In this study, we follow the isophotal PA and $\epsilon$ method described by Aguerri et al. (2009), Barazza et al. (2009), Consolandi (2016) and Erwin (2018). The PA and $\epsilon$ are estimated by ellipse fitting to the surface brightness distribution of galactic disks. The ellipse fitting method has been extensively used in both observational and simulational efforts on galactic bars (Michel-Dansac & Wozniak 2006; Hilmi et al. 2020; Zhao et al. 2020; Fragkoudi et al. 2021).

We retrieve the corrected frames of our sample galaxies from the SDSS Data Archive Server, which has been bias-subtracted, flat-fielded, and pixel-defect corrected. We then crop 500 × 500 pixel (195″×195″) FITS images centered on the sample galaxies. We manually remove non-target components like foreground stars, background galaxies, and other artifacts individually. This study uses only $r$-band galaxy isophotes due to their higher S/N compared to $u$-, $i$-, and $z$-bands. Also, the galaxy size and mass catalog by Simard et al. (2011) and Mendel et al. (2014) contain $g$- and $r$-band. We do not find any differences between bar measurements in $g$- and $r$-bands.

We start by ellipse fitting to the galaxy isophotes using the IDL `MPFITELLIPSE` procedure based on the non-linear least squares fitting `MPFIT` routine (Markwardt 2009). From outside a galaxy toward its center, this procedure automatically fits each contour line with the formula of an ellipse like

$$x(t) = a\cos(t)\cos\phi - b\sin(t)\sin\phi$$
$$y(t) = a\cos(t)\sin\phi + b\sin(t)\cos\phi, \quad (1)$$

where $a$ and $b$ are values of the major and minor axes, and $\phi$ is the PA i.e., the misalignment between the x-axis and the major axis of a fitted ellipse. By definition, $\epsilon$ for each fitted ellipse is calculated by

$$\epsilon = \sqrt{1 - (b/a)^2} \,. \quad (2)$$

The location where the bar ends in face-on galaxies produces a peak of the $\epsilon$ profile with a nearly constant PA within the bar radius (Wozniak et al. 1995; Jogee et al. 2004; Marinova & Jogee 2007; Barazza et al. 2008; Aguerri et al. 2009; Consolandi 2016). Thus, the bar ends are identified at the position where the $\epsilon$ is maximum and the PA begins to change rapidly. To identify barred galaxies automatically, we compare whether the $\epsilon$ profile exhibits a rapid increase followed by a rapid decrease ($\Delta\epsilon \geq 0.08$), and the PA is roughly constant within the bar radius ($\Delta$PA $\leq 40°$). Using a simple test on artificial galaxies, Aguerri et al. (2009) showed that these criteria maximize the bar identification with the lowest bad/spurious detection. Figure 1 illustrates the best examples of the radial profile of PA and $\epsilon$ for barred galaxies in our sample. The red vertical dashed line in the middle and right columns represents where the bar ends. The $\epsilon$ value changes more significantly with a lower measurement error than PA. Thus, we derive bar radius $r_{\mathrm{Bar}}$ as the radius where $\epsilon$ reaches the maximum, not based on the variation of PA.

Although we use nearly face-on galaxies, the inclination of the galactic disk still affects the measurement of the bar length and axis ratio. In order to conduct a more accurate analysis, we follow the correction formula for the projection effect by Martin (1995), such that

$$L_{\mathrm{Bar}} = 2\,r_{\mathrm{Bar}}\left(\cos^2\theta_a + \sec^2 i \sin^2\theta_a\right)^{1/2}, \quad (3)$$

where $L_{\mathrm{Bar}}$ is the de-projected bar length, $\theta_a$ is the misalignment angle between the major axis of the bar and the galactic disk, and $i$ is the inclination of the galaxy. To estimate the bar length, $r_{\mathrm{Bar}}$ is multiplied by two. The de-projected bar axis ratio $(b/a)_{\mathrm{Bar}}$ is also derived from the formula, such that

$$(b/a)_{\mathrm{Bar}} = \frac{b}{a}\left[\frac{\cos^2\theta_b + \sec^2 i \sin^2\theta_b}{\cos^2\theta_a + \sec^2 i \sin^2\theta_a}\right]^{1/2}, \quad (4)$$

where $\theta_a$ and $\theta_b$ are the misalignment angle between the major and minor axes of the bar and the galactic disk, respectively. This de-projection procedure is widely adopted by many observational studies of barred galaxies (e.g., Gadotti et al. 2007; Oh et al. 2012; Slavcheva-Mihova & Mihov 2011; Zou et al. 2014; Font et al. 2017; Tahmasebzadeh et al. 2021).

Figure 2 shows postage-stamp examples of the multi-band SDSS images of barred galaxies. We identify 3662 barred galaxies out of ∼10,000 nearly face-on disk galaxies in the local universe. The fraction of barred galaxies (∼30%) in our sample is consistent with other previous observations (Eskridge et al. 2000; Laurikainen & Salo 2002; Marinova & Jogee 2007; Sheth et al. 2008; Cameron et al. 2010; Masters et al. 2011; Hoyle et al. 2011; Skibba et al. 2012; Melvin et al. 2013; Melvin et al. 2014; Simmons et al. 2017). Figure 3 shows the distribution of resulting $L_{\rm Bar}$ and $(b/a)_{\rm Bar}$ of barred galaxies among our sample. On average, longer bars tend to be more elongated. This result is well consistent with previous observations (Martin 1995; Hoyle et al. 2011; Buta et al. 2015; Consolandi 2016; Lee et al. 2020).

### 2.3. Control Sample

The SF and AGN activity do not evolve independently, and they cooperate/compete with each other in barred galaxies (Gadotti & Eustáquio de Souza 2004; Lin et al. 2013; Robichaud et al. 2017; Khoperskov et al. 2018; Kim & Choi 2020; Kim et al. 2020). Robichaud et al. (2017) demonstrated that the AGN feedback generally pushes gas outward and inhibits further SF in unbarred galaxies; however, the effect of bar-driven gas inflow compensates for the negative AGN feedback in barred galaxies. This complicated interplay between SF and AGNs can diminish the imprints of bars on them. Many observations showed that positive and negative AGN feedback on SF happens simultaneously in barred galaxies (e.g., NGC 6810 by Strickland 2007; NGC 1097 by Lin et al. 2013; NGC 1068 by García-Burillo et al. 2014; NGC 1365 by Venturi et al. 2017; NGC 5728 by Shin et al. 2019; NGC 5643 by García-Bernete et al. 2021). Thus, it needs to be investigated how a bar plays a role in SF and AGN activity individually without the contamination by the interplay between themselves.

Previous studies clearly showed that the bar fraction depends on redshift (e.g., Sheth et al. 2008; Melvin et al. 2014; Simmons et al. 2014; Cisternas et al. 2015; Vera et al. 2016), stellar mass (e.g., Masters et al. 2011; Díaz-García et al. 2016; Vera et al. 2016; Erwin 2018), B/T (e.g., Vera et al. 2016; Kataria & Das 2018; Lee et al. 2019), and environments (e.g., Skibba et al. 2012; Lin et al. 2014; Vera et al. 2016; Sarkar et al. 2021; Smith et al. 2022; Tawfeek et al. 2022). Moreover, it is well established that the existence of AGNs also relies on the intrinsic properties of galaxies. Hence, we first classify barred galaxies into non-AGN and AGN-hosting galaxies based on the BPT diagnostic. We then carefully construct control samples of unbarred galaxies for non-AGN and AGN barred galaxies, separately. We examine the bar-driven effect on SF only for non-AGN barred galaxies, whereas AGN-hosting barred ones are selected to investigate links between bars and BHs only.

Figure 4 shows the distributions of redshift, stellar mass, B/T, and local density, Log($\Sigma_{45}$), of non-AGN and AGN barred galaxies with respect to unbarred galaxies. The projected local density is derived by

$$\Sigma_{\rm N} = \frac{N}{\pi d_{\rm N}^2}\,, \quad (5)$$

where $d_{\rm N}$ is the comoving distance to the $\rm N^{th}$ closest neighboring galaxies. Log($\Sigma_{45}$) is defined as the average of Log($\Sigma_{\rm N}$) for N = 4 and 5 as described by Baldry et al. (2006). The redshift distribution reveals that barred galaxies become less common at higher redshifts. In addition, analysis of other parameters shows that barred galaxies tend to have lower B/T, higher stellar mass, and higher Log($\Sigma_{45}$) compared to unbarred galaxies. This result is consistent with previous studies (Nair & Abraham 2010; Lee et al. 2012a; Melvin et al. 2014; Simmons et al. 2014; Lee et al. 2019). Both observations and simulations demonstrate that it is difficult to produce bars in gas-rich, high-redshift galaxies (Bournaud & Combes 2002; Villa-Vargas et al. 2010; Athanassoula et al. 2013; Seo et al. 2019; Zhou et al. 2020). Galactic interactions, which are more common in dense environments, are a major contributor to bar formation, as evidenced by the higher bar fraction among interacting galaxies in such environments (Skibba et al. 2012; Lee et al. 2012a; Lin et al. 2014; Sarkar et al. 2021).

To account for these intrinsic differences, we randomly select galaxies from the sample of unbarred galaxies to have a similar distribution of redshift (bin range of ± 0.005 dex), stellar mass (± 0.01 dex), B/T (± 0.01 dex), and Log($\Sigma_{45}$) (± 0.01 dex) for non-AGN and AGN barred galaxies. In Figure 5, we compare barred galaxies and the control sample. The distribution of the control sample is nearly identical to those of barred galaxies.

### 3. NEW BAR PARAMETERS

Conventionally, the relative bar size, which is normalized by the host galaxy's size, and $b/a$ of the bar are widely used to represent the bar strength. These parameters are assumed to be independent of stellar mass and B/T. However, there are still detectable dependencies of stellar and bulge mass on relative bar size even after the normalization (see the results of Hoyle et al. 2011; Oh et al. 2012; George & Subramanian 2021; Kim et al. 2021). Only a few efforts investigated the possibility of intrinsic dependencies on the bar size measurement. For example, Erwin (2019) used the $\rm S^4G$ bar identification method from Buta et al. (2015) and measurements by Herrera-Endoqui et al. (2015) to investigate the relationship between bar size and various parameters, including stellar mass and disk size. They claimed that barred galaxies tend to be more extended in size at the same stellar mass and there is a two-fold correlation between bar length and stellar mass.

The different efficiency of bar growth depending on galaxy mass can explain the correlation between relative bar size and stellar mass. Kraljic et al. (2012) found that the bar formation usually happens at redshift $z \sim 0.8 - 1$, and more massive galaxies ($M_*/M_\odot \sim 10^{11}$) form their bar structures earlier. Using the EAGLE simulation, Cavanagh et al. (2022) found that the episodes of creation, destruction, and resurgence of bar structures are repeated with a mean lifetime of $\sim 2.24$ Gyr, and some galaxies undergo multiple bar formations from $z = 1$ to 0. The lifetime for bars in lower mass galaxies ($M_*/M_\odot < 10^{10.5}$) is slightly shorter ($\sim 2.01$ Gyr) than bars in more massive galaxies ($\sim 2.46$ Gyr). Rosas-Guevara et al. (2022) identified $\sim 1,000$ barred galaxies using the TNG50 simulation and traced the evolution of their bars at the redshift range from $z = 4$ to 0. They showed that bar extents and disk scale lengths positively correlate with stellar mass, and the dependence on disk scale length is stronger. Thus, the relative ratio between bar extents and disk scale lengths is not a flat function of the disk scale length and shows a decreasing trend. In observations, using a sample of 257 galaxies between $0.1 < z \leq 0.84$, Sheth et al. (2012) found that the dynamically hotter disks, increased turbulence of gas accretion, and elevated interaction/merger rates hinder bar formation in lower mass galaxies specifically at higher redshift. Using a classification from the GZ 2, Skibba et al. (2012) showed that bars favor inhabiting more massive and redder haloes irrespective of environmental influences. These studies speculated that more massive galaxies construct bars earlier and have longer time-scales for their bars to grow in size. This finding suggests that it is important to consider intrinsic biases when studying bar properties. As a result, in this section, we compare how the measured bar properties vary with intrinsic galactic properties and attempt to derive new bar parameters that avoid stellar mass and B/T biases.

### 3.1. The Dependence of Bar Properties

In Figure 6, we show the distribution of measured bar properties on the stellar mass versus B/T plane. We use two bar parameters: corrected and normalized bar size, $L_{\rm Bar}/2R_{50}$, where $R_{50}$ is the radius which contains 50% of the Petrosian fluxes, and bar's axis ratio, $(b/a)_{\rm Bar}$. The bar properties are correlated with both stellar masses and B/T. Even when using the normalized bar length, $L_{\rm Bar}/2R_{50}$, the stellar mass and B/T dependencies are clearly detectable. Moreover, we find an interesting and unexpected discrepancy between non-AGN and AGN barred galaxies. For non-AGN barred galaxies, the bar length, $L_{\rm Bar}/2R_{50}$, positively correlates with the stellar mass. In contrast, for AGN barred galaxies, $L_{\rm Bar}/2R_{50}$ decreases as stellar mass increases. For non-AGN barred galaxies, $(b/a)_{\rm Bar}$ shows the complicated bimodal distribution. $(b/a)_{\rm Bar}$ decreases as stellar mass increases at the lower stellar mass range ($M_*/M_\odot < 10^{10.25}$), whereas $(b/a)_{\rm Bar}$ of massive galaxies ($M_*/M_\odot > 10^{10.25}$) are strongly correlated with B/T than stellar mass. In contrast, $(b/a)_{\rm Bar}$ of AGN barred galaxies exhibit a more monotonic trend on the parameter space. More massive and bulge dominant AGN barred galaxies show larger $(b/a)_{\rm Bar}$. The discrepancy between non-AGN and AGN barred galaxies suggests that bar-driven evolution may affect these types of galaxies differently, with varying degrees of efficiency.

In Figure 7 and Figure 8, we respectively compare the distribution of stellar mass and B/T of galaxies with different $L_{\rm Bar}/2R_{50}$ and $(b/a)_{\rm Bar}$. Many previous studies on barred galaxies typically treated bar length and axis ratio as being independent of stellar mass and B/T, and classified bars as weak or strong based on the bar's length or $b/a$ parameters only. Following the conventional definition, we divide stronger and weaker bars based on the medians of bar properties; $L_{\rm Bar}/2R_{50}$ = 1.0 and $(b/a)_{\rm Bar}$ = 0.28 lines. As illustrated in Figure 7, non-AGN barred galaxies with relatively longer bars ($L_{\rm Bar}/2R_{50} > 1.0$) are more massive and more bulge dominant than non-AGN galaxies with shorter bars, whereas AGN barred galaxies with longer bars are slightly less massive than AGN galaxies with shorter bars. This opposite correlation between stellar mass bias and $L_{\rm Bar}/2R_{50}$ for non-AGN and AGN barred galaxies is consistent with our result illustrated above. However, due to the shallower range of stellar mass for AGN barred galaxies, it is difficult to detect a strong difference in the stellar mass distribution between AGN barred galaxies with shorter and longer bars. On the other hand, the effect of division by $(b/a)_{\rm Bar}$ is more prominent for B/T distribution than by $L_{\rm Bar}/2R_{50}$. As illustrated in Figure 8, non-AGN and AGN barred galaxies with relatively more elongated bars ($(b/a)_{\rm Bar} < 0.28$) are less bulge dominant than galaxies with rounder bars. This implies that dividing stronger and weaker bars by $L_{\rm Bar}/2R_{50}$ and $(b/a)_{\rm Bar}$ is insufficient to eliminate bias related to stellar mass and B/T. It is well established that SF and AGN activities are tightly connected to those intrinsic values. Thus, it is necessary to derive new bar parameters independent of stellar mass and B/T.

### 3.2. *The Definition of New Bar Parameters*

To derive new parameters independent of the effect of stellar mass and B/T together, we compare the 3-D distribution of stellar mass, B/T, and bar properties. Then we define new parameters $\Delta\left[L_{\rm Bar}/2R_{50}\right]$ and $\Delta\left[(b/a)_{\rm Bar}\right]$ as the differences in $L_{\rm Bar}/2R_{50}$ and $(b/a)_{\rm Bar}$ with respect to values on the best-fitted polynomial planes which are defined by the formula, such that

$$\begin{aligned} z = a + bx + cy + dx^2 + exy + fy^2 + \\ gx^3 + hx^2y + ixy^2 + jy^3, \end{aligned} \quad (6)$$

where $x$ and $y$ are stellar mass and B/T, respectively, with the different sets of coefficients for non-AGN and AGN barred galaxies. Each value of coefficients is shown in Table 1. After the multi-parameter normalization, our newly defined bar parameters are nearly independent of the two host galaxy properties. Figure 9 is the same as Figure 6 but for $\Delta\left[L_{\rm Bar}/2R_{50}\right]$ and $\Delta\left[(b/a)_{\rm Bar}\right]$ with a zoomed-in ($\times 0.1$) range of the parameter values. Two parameters $\Delta\left[L_{\rm Bar}/2R_{50}\right]$ and $\Delta\left[(b/a)_{\rm Bar}\right]$ still rely on stellar mass and B/T, but the dependencies are fairly negligible compared to $L_{\rm Bar}/2R_{50}$ and $(b/a)_{\rm Bar}$, which have been conventionally used in previous studies.

## 4. INTERPLAY OF BARS WITH SF AND AGNS

The impact of bars on the evolution of SF and AGN activities is not yet fully understood, despite the importance suggested by numerous studies. Some researches showed that galaxies with longer bars exhibit enhanced SFRs, whereas others suggested that bars halt further SF. Similarly, whether the existence of bars makes more vigorous BH activity is inconclusive. To find out how bars affect the evolution of galaxies, we delve into the sSFR and BH mass as functions of the bar properties.

### 4.1. *Interplay between Bars and SF Activities for Non-AGN Galaxies*

In the left panel of Figure 10, we show sSFR$_{\rm Global}$ of barred and unbarred galaxies as a function of stellar mass. We use non-AGN barred galaxies and their corresponding control sample to avoid contamination by AGN emissions. In general, barred galaxies are located on average ∼0.12 dex above the star-forming main sequence of unbarred galaxies. We define the residual of sSFR$_{\rm Global}$, $\Delta\mathrm{Log}(\mathrm{sSFR}_{\rm Global})$, as the difference in sSFR$_{\rm Global}$ between the barred galaxy sample and its corresponding unbarred control sample. The right panel illustrates the distribution of $\Delta$sSFR$_{\rm Global}$ on the parameter space of $\Delta\left[L_{\rm Bar}/2R_{50}\right]$ and $\Delta\left[(b/a)_{\rm Bar}\right]$. $\Delta\mathrm{Log}(\mathrm{sSFR}_{\rm Global})$ mainly correlates with the bar length parameter but not with the $b/a$ parameter. Non-AGN barred galaxies with longer bars tend to exhibit enhanced sSFR$_{\rm Global}$. However, the $b/a$ parameter does not show significant trends with SF enhancements.

Non-AGN barred galaxies exhibit a higher sSFR$_{\rm Fiber}$ compared to unbarred control galaxies in the left histogram of Figure 11. This is consistent with previous studies which suggested that the majority of bar-driven SF enhancement happens at the galactic center (Athanassoula 1994; Sakamoto et al. 1999; Sheth et al. 2005; Ellison et al. 2011; Coelho & Gadotti 2011; Wang et al. 2012; Zhou et al. 2015; Chown et al. 2019; Lin et al. 2020). We also fit Gaussian to the sSFR$_{\rm Fiber}$ distribution of barred galaxies, and find that the Gaussian peak locates at $\mathrm{Log}(\mathrm{sSFR}_{\rm Fiber}) = -9.93\,[\mathrm{yr}^{-1}]$. Then, following the definition from Kim et al. (2018), we define galaxies with $\mathrm{Log}(\mathrm{sSFR}_{\rm Fiber}) > -9.93\,[\mathrm{yr}^{-1}]$ as central starburst galaxies, and compare the number fraction of starburst galaxies, $f_{\rm SB}$, as functions of bar properties in the middle and right panels of Figure 11. We observe a significant positive correlation between the $\Delta\left[L_{\rm Bar}/2R_{50}\right]$ and $f_{\rm SB}$, such that central starburst galaxies are more commonly found in barred galaxies with longer bars. However, we do not observe any detectable trends in the $b/a$ parameters with the concentration of SF enhancements.

Our result is consistent with previous observations (Hawarden et al. 1986; Arsenault 1989; Martin 1995; Martinet & Friedli 1997; Ho et al. 1997; Alonso-Herrero & Knapen 2001; Hunt et al. 2008; Ellison et al. 2011; Coelho & Gadotti 2011; Lin et al. 2017; Chown et al. 2019; Lin et al. 2020) and simulations (Knapen et al. 1995; Sakamoto et al. 1999; Lin et al. 2013; Cole et al. 2014; Spinoso et al. 2017; Zana et al. 2019), which provided evidence that bar-induced SF correlates with bar's length only but not with bar's ellipticity. Lin et al. (2017) investigated central SF histories of 57 face-on disk galaxies through the CALIFA survey. They identified 17 'turnover galaxies' that show a rapid upturn of SF toward the center and found that the majority (15/17) of 'turnover galaxies' are barred. There are weak and positive correlations between the bar's physical length and the size of the SF turnover region, whereas there is no clear correlation with the ellipticity of bars. More recently, using 2D maps of EW(Hα), EW(Hδ$_{\rm A}$), and D$_{\rm n}$(4000) through the MaNGA survey, Lin et al. (2020) revealed that 89% of galaxies with centrally concentrated SF have bars. The size of the central star-forming region shows a tight correlation with one-third of the bar's physical length. They suggested that bars play an *"indispensable role"* in the SF enhancement. These stud-ies proposed that the presence of bars can stimulate the inflow of gas, enhancing the concentration and triggering SF at the center as the bar grows. This bar-driven evolution of SF results in a positive correlation between the length of the bar and the increase in central SF. In other words, bars in star-forming galaxies tend to become longer over time.

In contrast, other simulations (Spinoso et al. 2017; Khoperskov et al. 2018) claimed that strongly barred galaxies exhibit regulated SF activity. They suggested that the rapid SF driven by bars in the past could deplete the gas supply, leading to a decrease in the current SFR. Some observations support this bar-quenching scenario (Gavazzi et al. 2015; George et al. 2019b; Newnham et al. 2020; Fraser-McKelvie et al. 2020a; Fraser-McKelvie et al. 2020b; George et al. 2020; Géron et al. 2021; George & Subramanian 2021); however, stellar mass biased dependence remains in their results. For example, Gavazzi et al. (2015) found an increase of bar fraction with a decrease of SFRs for massive galaxies above specific threshold mass ($M_{\mathrm{Knee}}/M_\odot \sim 10^{9.5}$). Using a sample of 684 barred galaxies from the MaNGA survey, Fraser-McKelvie et al. (2020a) demonstrated that the effect of bars on SF is predominantly governed by stellar mass. Low-mass galaxies exhibit increased SF along with their bar structures, whereas massive galaxies show regulated SF. Géron et al. (2021) identified strongly and weakly barred galaxies from 1867 galaxies through a visual classification by Galaxy Zoo DECaLS (GZD). They concluded that strong bars are more prevalent in quiescent galaxies than in star-forming ones, but did not examine the effect of stellar mass bias. Intriguingly, however, when dividing their sample into star-forming and quiescent galaxies, only star-forming barred galaxies exhibit a positive correlation between the fiber SFR and bar length.

### 4.2. Interplay between Bars and BH Activities for AGN Galaxies

In Figure 12, we classify AGN-hosting galaxies based on the distribution on the BPT diagram and compare the number fraction of AGNs, $f_{\mathrm{AGN}}$, for barred galaxies and the unbarred control sample as a function of stellar mass. Both barred and unbarred galaxies show that $f_{\mathrm{AGN}}$ increases as stellar mass increases. However, there is no difference between barred and unbarred galaxies. Barred galaxies among our sample do not show a $f_{\mathrm{AGN}}$ enhancement with respect to unbarred ones.

To examine the bar-driven effect on the strength of AGN activity, we now estimate the BH mass by employing the well-established $M_{\mathrm{BH}}$–$\sigma_e$ relation (Novak et al. 2006), such that

$$\mathrm{Log}(M_{\mathrm{BH}}/M_\odot) = \alpha + \beta \, \mathrm{Log}(\sigma_e/200\,\mathrm{km\,s^{-1}}) \,, \quad (7)$$

where $M_{\mathrm{BH}}/M_\odot$ is the BH mass in solar mass units, $\sigma_e$ is the velocity dispersion of galaxies inside the effective radius in [km s$^{-1}$] units, and $\alpha$ and $\beta$ are regression coefficients. Due to the 1.5″ radius of the SDSS fiber coverage, we apply a correction for the velocity dispersion from Cappellari et al. (2006), such that

$$(\sigma_e/\sigma) = (R/R_e)^{(0.066 \pm 0.035)} \,, \quad (8)$$

where $\sigma$ and $\sigma_e$ are the measured and corrected velocity dispersion, $R$ is the radius encompassed by the SDSS 1.5″ fiber, and $R_e$ is the effective radius of galaxies. Many observations reported distinct offsets between barred and unbarred galaxies from the $M_{\mathrm{BH}}$–$\sigma_e$ relation (e.g., Gültekin et al. 2009; Graham & Li 2009; Graham et al. 2011; Sahu et al. 2019). Through collisionless N-body simulations, Brown et al. (2013) confirmed that orbital properties of stars in bars increase $\sigma_e$ by roughly 4 – 8 %. Table 2 gives examples of the regression coefficient of the $M_{\mathrm{BH}}$–$\sigma_e$ relation for all, barred and unbarred galaxies from the literature. Following these results, we adopted different coefficients for barred and unbarred galaxies. There are no significant differences between the results of each set of coefficients, thus we select the most recent value from Sahu et al. (2019).

The left panel of Figure 13 shows $M_{\mathrm{BH}}$ of AGN galaxies as a function of stellar mass. Intriguingly, when we adopt distinct $M_{\mathrm{BH}}$–$\sigma_e$ relations for barred and unbarred galaxies, barred galaxies have ~0.4 dex less massive BHs than unbarred ones at all stellar mass range. This is consistent with previous studies showing that BHs in barred galaxies are of around half the mass of BHs in unbarred galaxies for the same velocity dispersion, (Graham 2008; Graham & Li 2009; Graham 2014; Mutlu-Pakdil et al. 2018) and relatively less massive BHs are more common in barred galaxies (Oh et al. 2012). We estimate the residuals of $M_{\mathrm{BH}}$, $\Delta\mathrm{Log}(M_{\mathrm{BH}}/M_*)$, as the difference between $M_{\mathrm{BH}}$ of barred galaxies and $M_{\mathrm{BH}}$ of their corresponding control sample. The right panel shows the distribution of $\Delta\mathrm{Log}(M_{\mathrm{BH}}/M_*)$ on the parameter space of $\Delta[L_{\mathrm{Bar}}/2R_{50}]$ and $\Delta[(b/a)_{\mathrm{Bar}}]$. Galaxies with rounder bars tend to have more massive BHs. Specifically, galaxies with relatively rounder bars ($\Delta[(b/a)_{\mathrm{Bar}}] > 0.00$) show a significant correlation between $\Delta\mathrm{Log}(M_{\mathrm{BH}}/M_*)$ and $b/a$ parameters only. For a comparison between bars and AGN feeding processes, we utilize the $L_{\mathrm{[OIII]}}$ and accretion rate parameter, $\Re$, as a proxy for the BH accretion. Adopting the definition of $\Re$ from Heckman et al. (2004), where $\Re = \text{Log}(L_{[\text{OIII}]}/M_{\text{BH}})$, we find that barred AGN galaxies are ∼0.4 dex higher in $\Re$ than the unbarred control sample. A correlation with $\Re$ is observed only for $b/a$ parameters, indicating that galaxies with rounder bars have higher $\Re$ values. Since $\Re$ is directly linked to the value of BH mass, we do not include the results of $\Re$ in this paper.

Our result of the absence of an increase in the $f_{\text{AGN}}$ for barred galaxies over unbarred ones is consistent with the recent observations that bars do not yield AGNs (Hunt & Malkan 1999; Gadotti 2008; Zhang et al. 2009; Lee et al. 2012a; Cisternas et al. 2013; Galloway et al. 2015; Cisternas et al. 2015; Goulding et al. 2017). However, we show that AGN barred galaxies have lower BH mass than unbarred counterparts. Interestingly, unlike the results of SF enhancement, the $b/a$ parameter correlates more strongly with the residuals of BH mass: $\Delta M_{\text{BH}}$ increases for rounder bars. These results suggest that the bar does not directly induce AGNs, but the evolution of BHs can regulate bars' shape. Contrary to earlier studies proposing that bar-induced gas inflow can trigger AGN activity (Combes 2001; Emsellem et al. 2001; Alonso et al. 2013; Alonso et al. 2014; Emsellem et al. 2015; Fanali et al. 2015; Cisternas et al. 2015; Alonso et al. 2018; Audibert et al. 2019), our results suggest that the relationship between BHs and bars may be more complex, with BHs possibly having an influence on the bar evolution rather than the other way around.

Bars are self-regulated structures and can be destroyed by the growth of central mass concentration (CMC) and BHs (Norman et al. 1985; Hasan & Norman 1990; Pfenniger & Norman 1990; Hasan et al. 1993; Norman et al. 1996; Hozumi & Hernquist 1998; Combes 2000; Bournaud & Combes 2002 Das et al. 2003; Shen & Sellwood 2004; Athanassoula et al. 2005; Bournaud et al. 2005; Hozumi & Hernquist 2008; Hozumi 2012; Li et al. 2017; Kataria et al. 2020). Some observations also support the bar destruction scenario. For example, Mulchaey & Regan (1997) found that the bar fraction in Seyfert and normal galaxies are similar. To explain the absence of a link between the presence of bars and Seyfert nuclei, they proposed that most bars in unbarred Seyfert galaxies were recently destroyed along with the growth of BHs. Cheung et al. (2013) revealed distinct trends of bars in galaxies with disky pseudobulges ($n < 2.5$) and classical bulges ($n > 4$). They suggested a scenario in which the fate of bars relies on the existence of CMCs. The CMC in gas-poor galaxies can interrupt the formation of new bars and destroys pre-existing bars. Du et al. (2017) and Guo et al. (2020) also reported that similar bar destruction and limited BH growth are linked, even for the evolution of inner bars. When BH mass reaches ∼0.2 % of total stellar mass, inner bars embedded in large-scale bars begin to be destroyed. The destruction of inner bars halts bar-driven gas inflow further, and finally, BHs stop growing. This interplay between bar dissolution and BH fueling can explain our result of the absence of increased AGN fraction and the restricted distribution of BH mass for barred galaxies.

Bar ellipticity can be a good indicator to trace the process of bar destruction. Hasan & Norman (1990) and Hozumi (2012) simulated the evolution of a bar's shape during the BH growth. They found that as bars undergo destruction, bars become shorter and thicker. Additionally, their simulation showed that bars with a rounder shape are easier to be destroyed compared to elongated bars. Contrary to the bar length parameters, only a few observational studies have investigated bar ellipticity as an indicator of bar strength (e.g., Gadotti 2008; Cisternas et al. 2013; Díaz-García et al. 2016; Silva-Lima et al. 2022). Das et al. (2003) introduced bar ellipticity as the best indicator for the accumulating CMC, which leads to bar destruction. Using kinematics of cold gas of 29 barred galaxies based on their CO rotation curves through the Berkeley-Illinois-Maryland Association (BIMA) Survey of Nearby Galaxies (SONG), the authors found a clear correlation between bar ellipticity and CMCs. They showed that bars become rounder as CMCs grow, and ultimately bars are destroyed. This result is consistent with our findings that only bars' $b/a$ parameter has a positive correlation with the BH mass, whereas bar lengths do not show a significant correlation.

Previous studies suggested that the correlation between the presence of bars and the enhancement of AGN may be due to a selection bias, as both tend to be more common in more massive galaxies. This could create the impression that bars cause an increase in AGN activity when, in fact, the relationship may be a result of both being correlated with stellar mass. For example, Lee et al. (2012b) discovered that the AGN fraction is around two times higher in strongly barred galaxies (∼34.5%) than in unbarred ones (∼15.0%); however, the excess of the AGN incidence in barred galaxies diminishes when they compare barred and unbarred galaxies at fixed $u-r$ colors and stellar masses. Many other observations reported similar results: barred galaxies tend to host AGNs, but the bar-driven AGN enhancement usually happens in more massive galaxies (e.g., Oh et al. 2012; Alonso et al. 2013; Galloway et al. 2015; Silva-Lima et al. 2022). This is the main reason why we define new bar properties independent of stellar mass to avoid bias.

## 5. SUMMARY AND CONCLUSIONS

We investigated the effect of bars on SF and AGN activity using a large sample of barred galaxies from the SDSS DR12. Through our automatic bar measurement scheme, we identified 3662 barred galaxies out of ∼10,000 nearly face-on disk galaxies in the local universe ($0.02 < z < 0.06$). We estimated the length and axis ratio of bar structures taking into account the projection effect. We classified our sample into non-AGN and AGN barred galaxies based on the BPT diagram. To avoid sampling bias, we carefully constructed control samples of unbarred galaxies for non-AGN and AGN barred galaxies separately. Then, we compared the SF and AGN activity between barred and unbarred galaxies and how the behaviors depend on bar parameters.

Our main results are summarised as follows.

1. Our measured bar properties, including bar's physical size and $b/a$ parameter, are consistent with previous observations. Longer bars tend to be more elongated. Both $L_{\rm Bar}$ and $(b/a)_{\rm Bar}$ depend on stellar mass and B/T.

2. Even after the bar size is normalized by the galaxy size itself, the relative bar size, $L_{\rm Bar}/2R_{50}$ and axis ratio of bar, $(b/a)_{\rm Bar}$, still show detectable correlations with stellar mass and B/T. Moreover, non-AGN and AGN barred galaxies exhibit distinct dependencies of intrinsic values. To avoid the biases by intrinsic dependencies, we define new bar parameters, $\Delta\,[L_{\rm Bar}/2R_{50}]$ and $\Delta\,[(b/a)_{\rm Bar}]$, using the 3-D distribution of stellar mass, B/T, and conventional bar properties.

3. Non-AGN barred galaxies are located ∼0.12 dex above the star-forming main sequence of the corresponding unbarred control sample in the stellar mass versus $\rm sSFR_{Global}$ plane. Galaxies with longer bars are more likely to exhibit central starburst activity. The strength of SF activity of barred galaxies depends on the bar length parameter, but not on the $b/a$ parameter. The $\Delta\rm Log(sSFR_{Global})$ and $f_{\rm SB}$ increase as the bar length increases.

4. AGN barred galaxies have ∼0.4 dex less massive BHs than the corresponding unbarred control sample. Contrary to the results of SF enhancement, the residuals of BH mass between barred and unbarred galaxies primarily depend on the $b/a$ parameter, rather than on the bar length parameter. $\Delta\rm Log(M_{BH}/M_*)$ increases for rounder and shorter bars.

Our results imply that bars are linked to SF and AGN activities in different ways. As bars grow in length, the gas inflow through the bar triggers higher sSFR. Thus, non-AGN barred galaxies generally exhibit more induced sSFR than unbarred ones, and SF enhancement is usually correlated with the bar length parameter. The existence of bars itself does not influence the presence of AGNs. However, bars hinder the mass growth of BHs, and barred AGN galaxies have limited BH mass distribution more than unbarred ones. Contrary to the result of SF enhancement, BH mass more significantly depends on the bar's $b/a$ parameters rather than bar lengths. This result can be explained by bar destruction during BH growth in barred galaxies which is supported by previous simulations. In a nutshell, bars in non-AGN galaxies are growing, whereas bars in AGN galaxies are under dissolution. This scenario is consistent with the result of the opposite direction of stellar mass bias for non-AGN and AGN barred galaxies. A visual demonstration of chronological bar-driven evolution on SF and BH activities is presented in Figure 14. Following these results, for the first time, we suggest that the effect of bar-driven evolution on SF and BH activities should be traced by different bar parameters, bar's length, and $b/a$, respectively.

## ACKNOWLEDGMENTS

S.-J.Y. acknowledges support from the Mid-career Researcher Program (No. 2019R1A2C3006242) and the Basic Science Research Program (No. 2022R1A6A1A03053472) through the National Research Foundation (NRF) of Korea. S.P. acknowledges support from the Mid-career Researcher Program (No. RS-2023-00208957) through the NRF of Korea. Funding for the Sloan Digital Sky Survey (SDSS) has been provided by the Alfred P. Sloan Foundation, the Participating Institutions, the National Aeronautics and Space Administration, the National Science Foundation, the U.S. Department of Energy, the Japanese Monbukagakusho, and the Max Planck Society. The SDSS Web site is http://www.sdss.org/. The SDSS is managed by the Astrophysical Research Consortium (ARC) for the Participating Institutions. The Participating Institutions are The University of Chicago, Fermilab, the Institute for Advanced Study, the Japan Participation Group, The Johns Hopkins University, Los Alamos National Laboratory, the Max-Planck-Institute for Astronomy (MPIA), the Max-Planck-Institute for Astrophysics (MPA), New Mexico State University, University of Pittsburgh, Princeton University, the United States Naval Observatory, and the University of Washington.

Joint Evolution of Bars, SF, and AGNs          15

**Table 1.** The values of coefficients of best-fitted polynomial planes for 3-D distribution of stellar mass, B/T and bar properties including $L_{\rm Bar}/2R_{50}$ and $(b/a)_{\rm Bar}$.

| Galaxy types | Bar properties | $a$ | $b$ | $c$ | $d$ | $e$ | $f$ | $g$ | $h$ | $i$ | $j$ |
|---|---|---|---|---|---|---|---|---|---|---|---|
| Non-AGN barred | $L_{\rm Bar}/2R_{50}$ | -131 | 42.43 | -116 | $-4.56$ | 24.38 | $-31.98$ | 0.163 | $-1.29$ | 3.87 | $-6.49$ |
| | $(b/a)_{\rm Bar}$ | 14.18 | $-4.43$ | 11.27 | 0.47 | $-2.04$ | $-5.34$ | $-0.016$ | 0.0912 | 0.49 | 0.60 |
| AGN barred | $L_{\rm Bar}/2R_{50}$ | 367.1 | $-105.5$ | 55.85 | 10.14 | $-11.58$ | 18.56 | $-0.33$ | 0.60 | $-1.61$ | $-1.21$ |
| | $(b/a)_{\rm Bar}$ | 0.43 | 0.65 | $-34.16$ | $-0.13$ | 6.33 | 1.40 | 0.0061 | $-0.29$ | $-0.18$ | 0.45 |

**Table 2.** The values of regression coefficients of the $M_{\rm BH}$–$\sigma_e$ relation, $\mathrm{Log}(M_{\rm BH}/M_\odot) = \alpha + \beta\,\mathrm{Log}(\sigma_e/200\,\mathrm{km\,s^{-1}})$, from literature.

| References | Morphological aspects | $\alpha$ | $\beta$ |
|---|---|---|---|
| Gültekin et al. 2009 | Without distinction | $8.12 \pm 0.08$ | $4.24 \pm 0.41$ |
| | Barred | $7.67 \pm 0.115$ | $4.08 \pm 0.751$ |
| | Unbarred | $8.19 \pm 0.087$ | $4.21 \pm 0.446$ |
| Graham et al. 2011 | Without distinction | $8.13 \pm 0.005$ | $5.13 \pm 0.34$ |
| | Barred | $7.80 \pm 0.10$ | $4.34 \pm 0.56$ |
| | Unbarred | $8.25 \pm 0.06$ | $4.57 \pm 0.35$ |
| Sahu et al. 2019 | Without distinction | $8.04 \pm 0.12$ | $4.47 \pm 0.80$ |
| | Barred | $8.01 \pm 0.10$ | $4.05 \pm 0.54$ |
| | Unbarred | $8.36 \pm 0.06$ | $5.46 \pm 0.34$ |

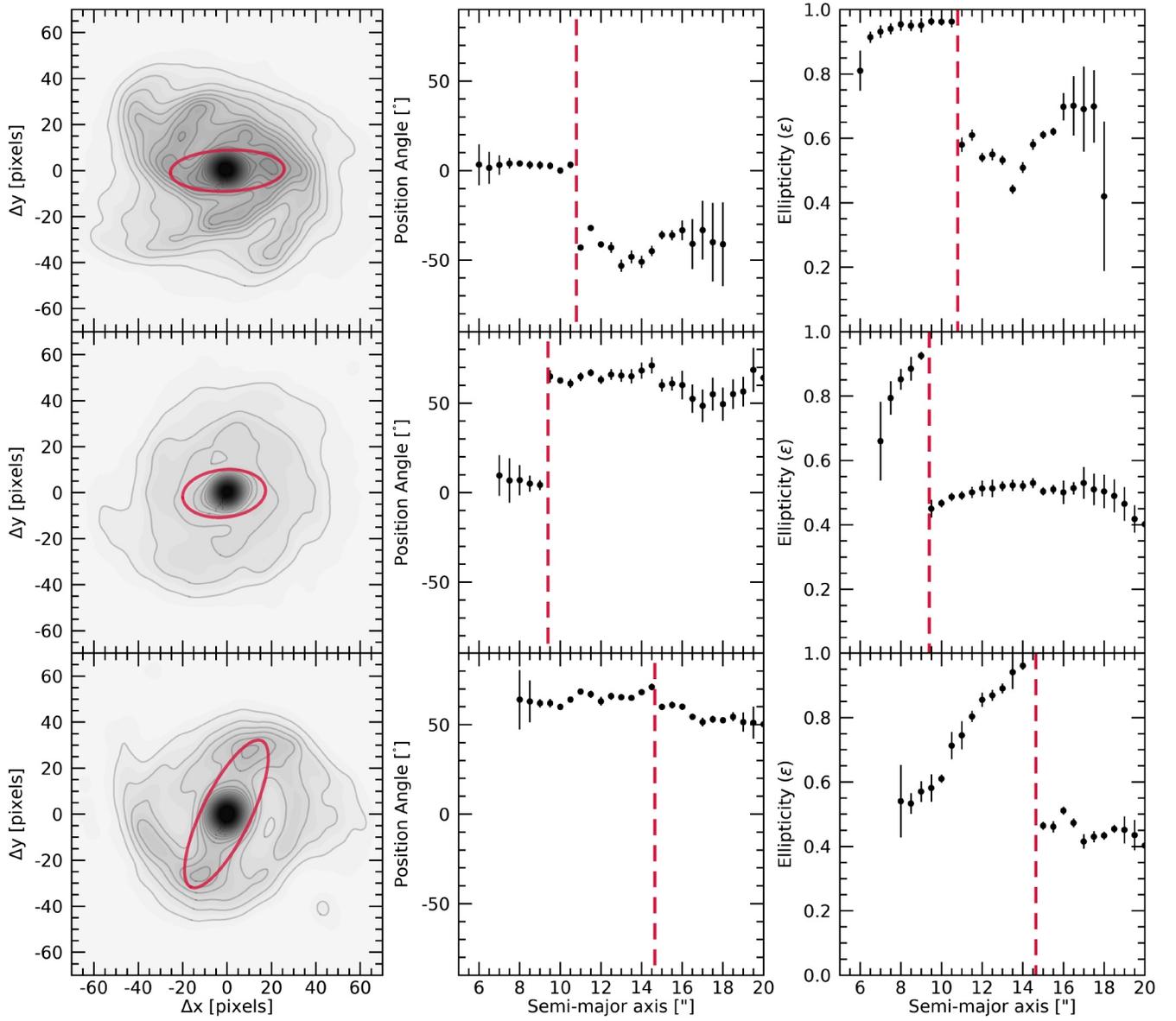

**Figure 1.** Examples of bar identification and measurements. (Left column) The identified bar structures are illustrated by a red ellipse on the contour image of each galaxy. (Middle) The radial profiles of the positional angle of isophotes through ellipse fitting from the center to the outskirt. The vertical red dashed line represents where the bar ends. (Right) The same as the left column but for bar ellipticity.

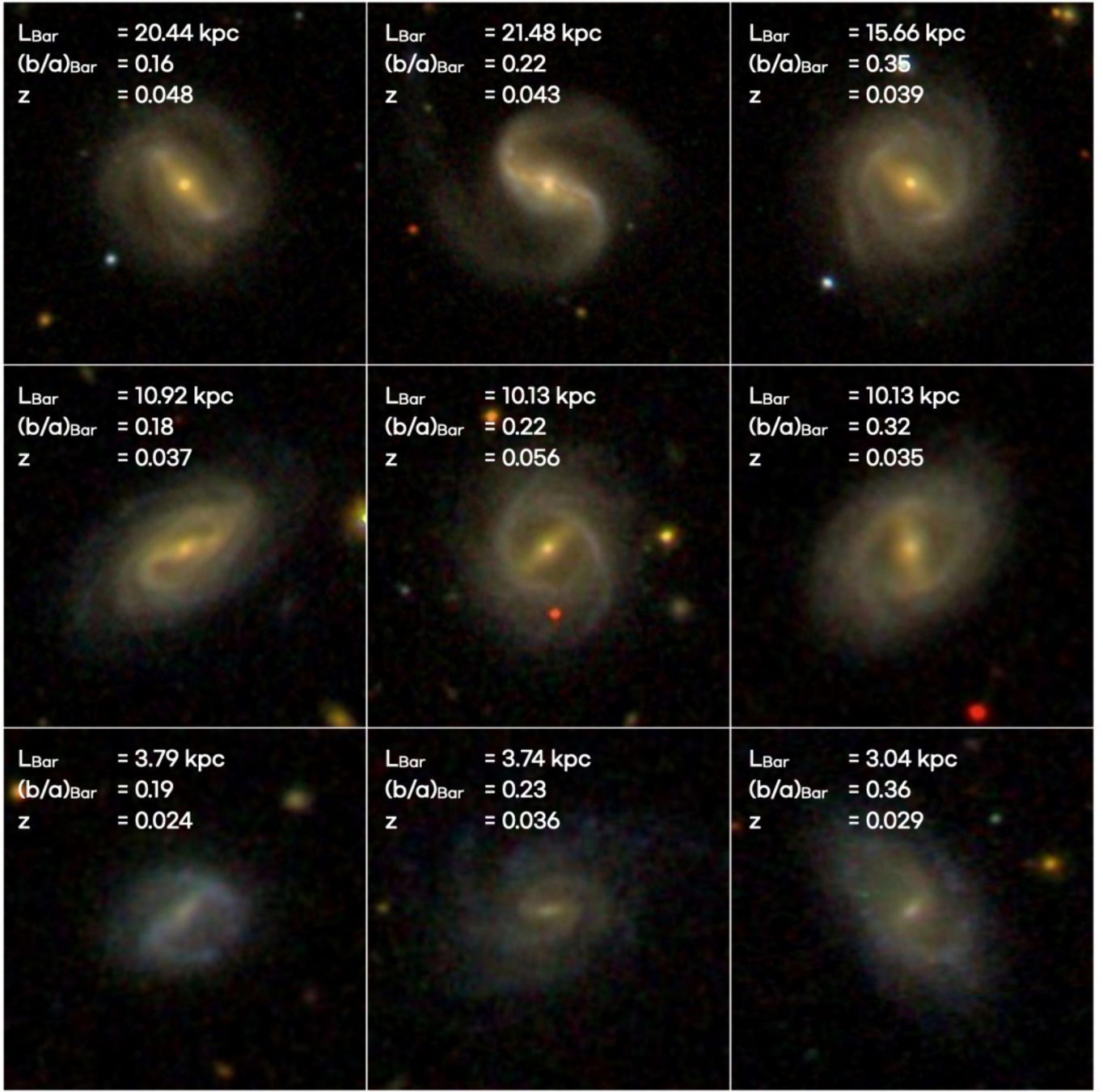

**Figure 2.** An example of multi-band SDSS images of barred galaxies that are identified using our automatic bar classification and measurement scheme. For comparison purpose, the measured bar length, bar axis ratio, and redshift of each galaxy are shown. The bar length increases from bottom to top, and the bar axis ratio increases from left to right.

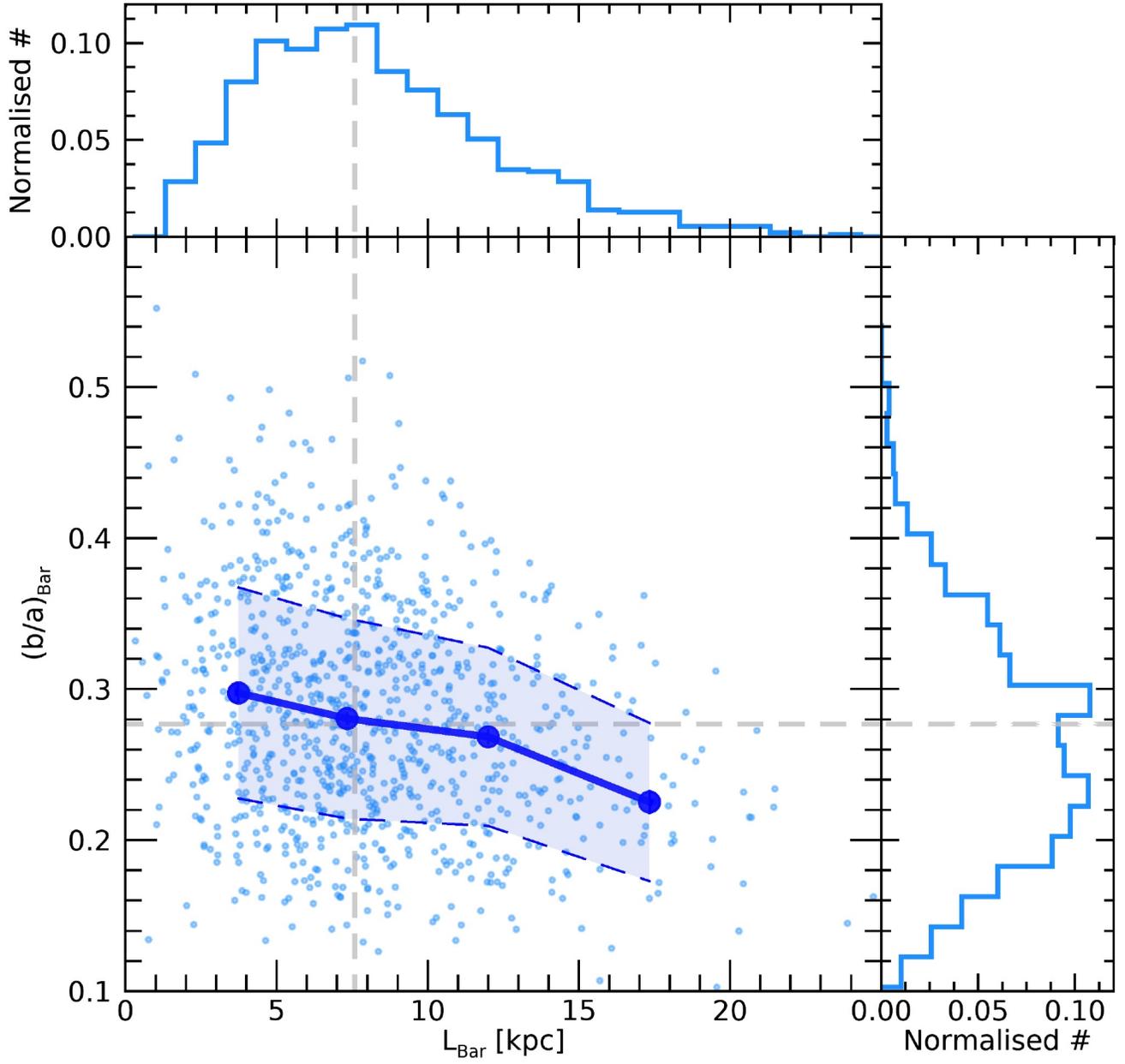

**Figure 3.** The distribution of measured bar's physical size, $L_{\rm Bar}$, and the projected bar axis ratio, $(b/a)_{\rm Bar}$, properties. Blue dots and line show the median distribution of two properties. Light blue region shows standard deviation. The grey dashed lines show the location of median values. The upper and right panels show histogram of $L_{\rm Bar}$ and $(b/a)_{\rm Bar}$, respectively.

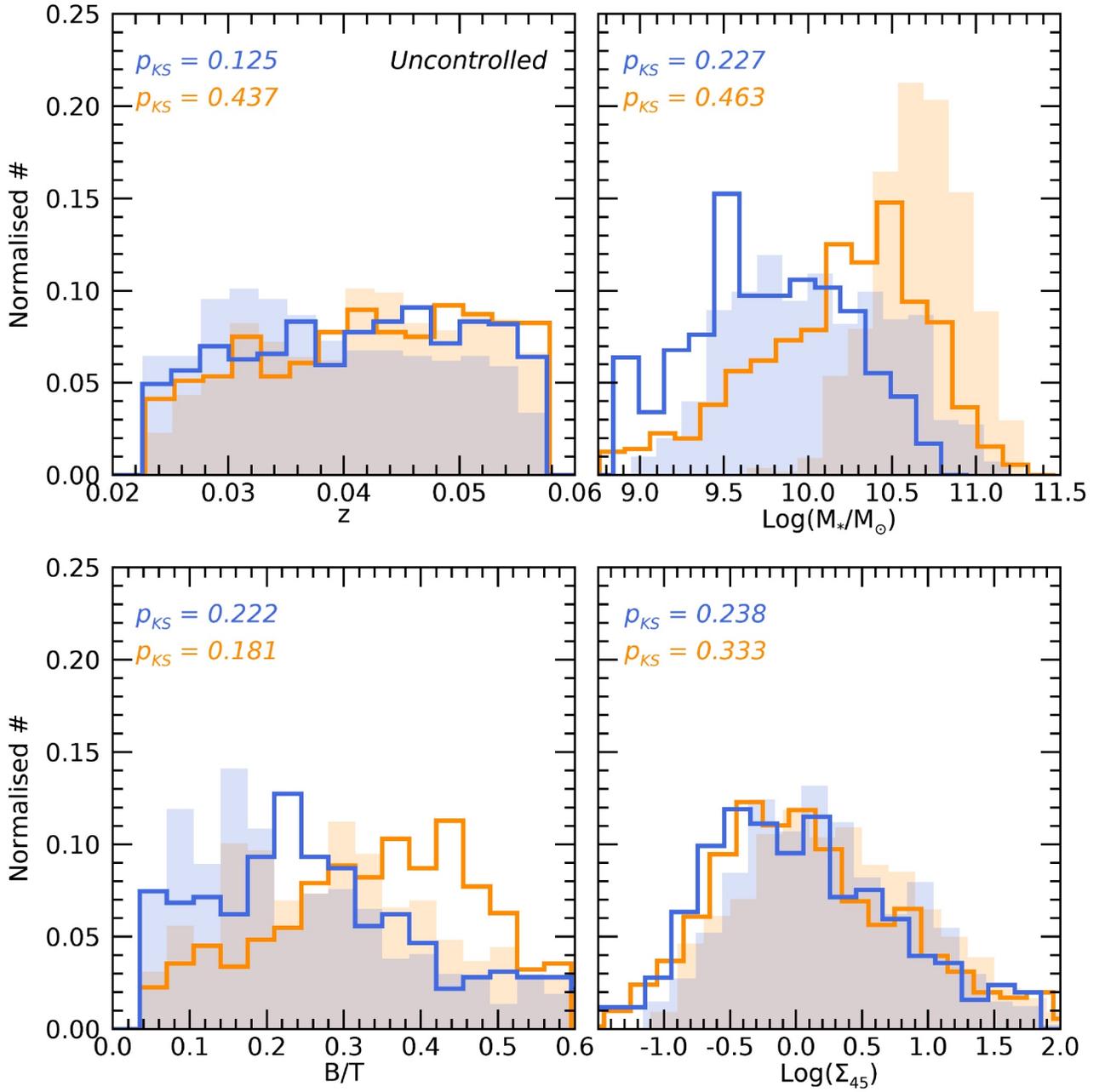

**Figure 4.** The distribution of the redshift (upper left), stellar mass (upper right), B/T (lower left), and local density (lower right) for unbarred (empty histograms depicted by thick lines) and barred galaxies (filled histogram). Non-AGN and AGN galaxies are represented by blue and orange colors, respectively. For each normalized histogram, the integral under the histogram is equal to one. Each p-value by the KS test is given on each panel.

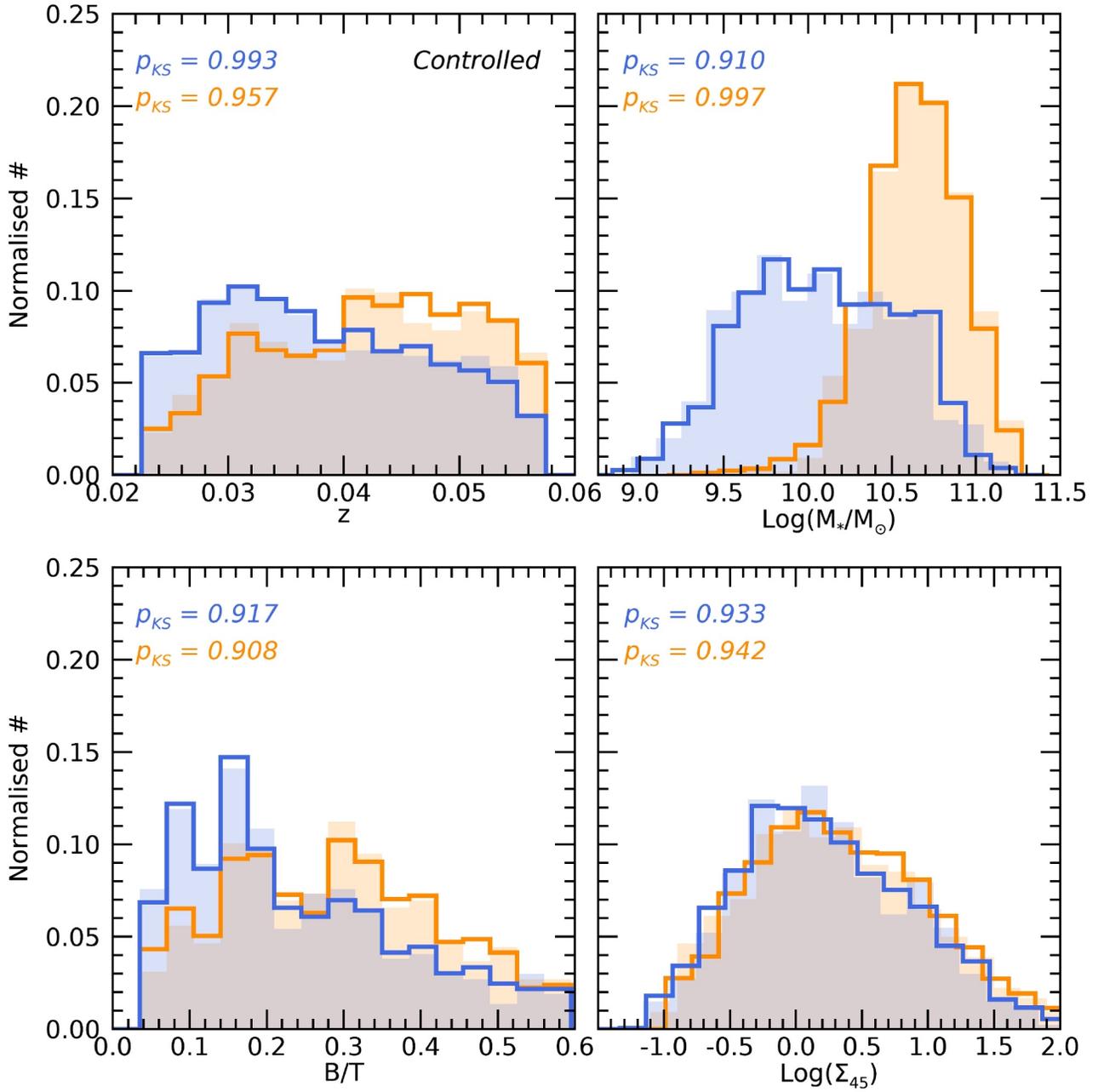

**Figure 5.** The same as Figure 4 but for the controlled unbarred sample.

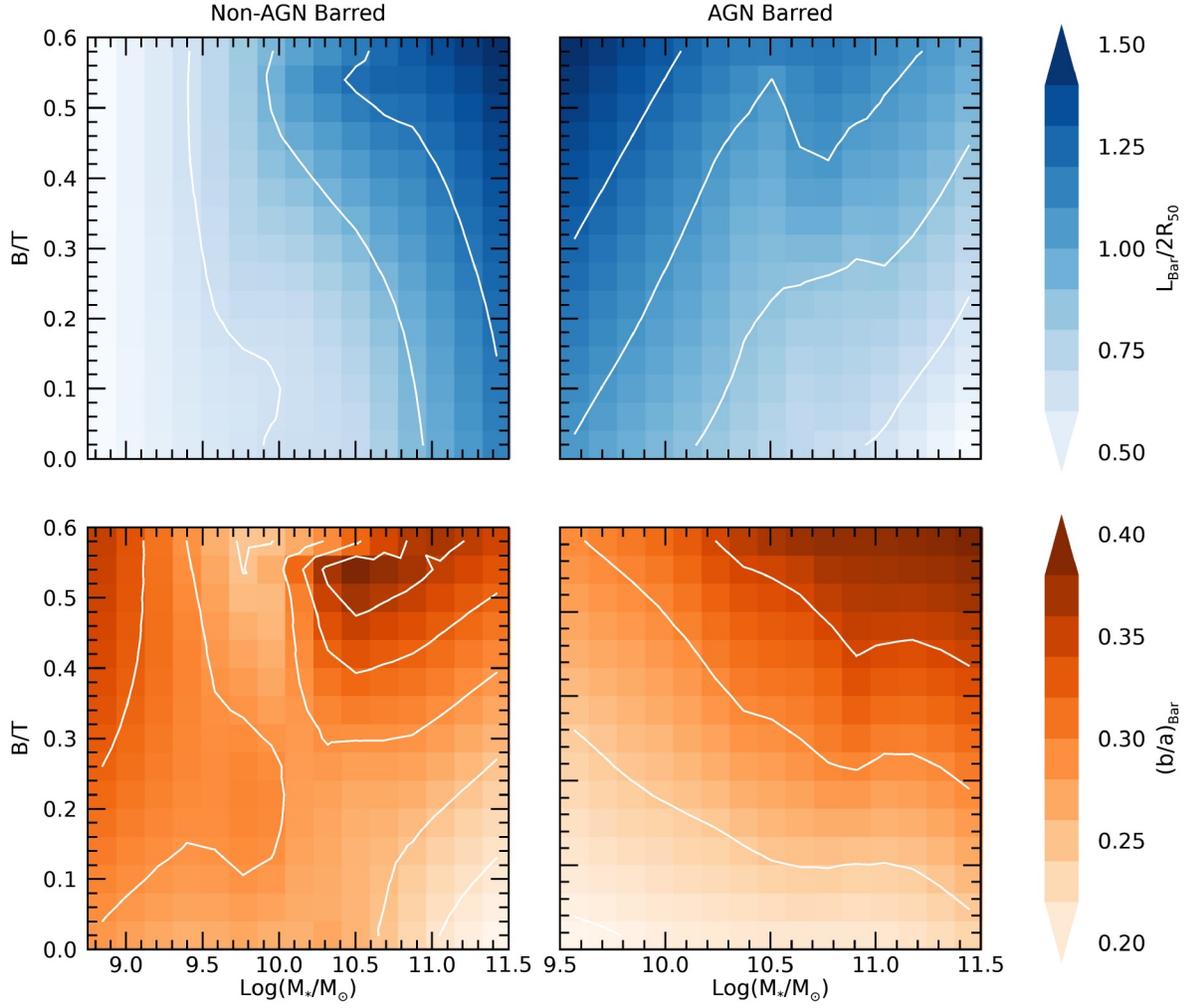

**Figure 6.** Upper panels: The 2D histogram of relative bar size, $L_{\rm Bar}/2R_{50}$, on the stellar mass and B/T parameter space for non-AGN (left) and AGN (right) barred galaxies, respectively. The local regression scheme of Cappellari et al. (2013) is applied. The bluer color represents relatively longer bars with larger $L_{\rm Bar}/2R_{50}$ as shown in the color bar on the right side. For comparison, the $0.5\sigma$, $1\sigma$, $1.5\sigma$, and $2\sigma$ contours are also shown as white lines. Please note that AGNs inhabit more massive galaxies; thus, the range of the x-axis is different for non-AGN and AGN galaxies. For comparison, median values and the range of standard deviations are shown above each histogram. Lower panels: The same as upper panels but for the distribution of $(b/a)_{\rm Bar}$. The redder color represents rounder bars with larger $(b/a)_{\rm Bar}$. Please note that AGN barred galaxies are distributed in the range of relatively more massive stellar mass ($M_*/M_\odot \geq 10^{9.5}$).

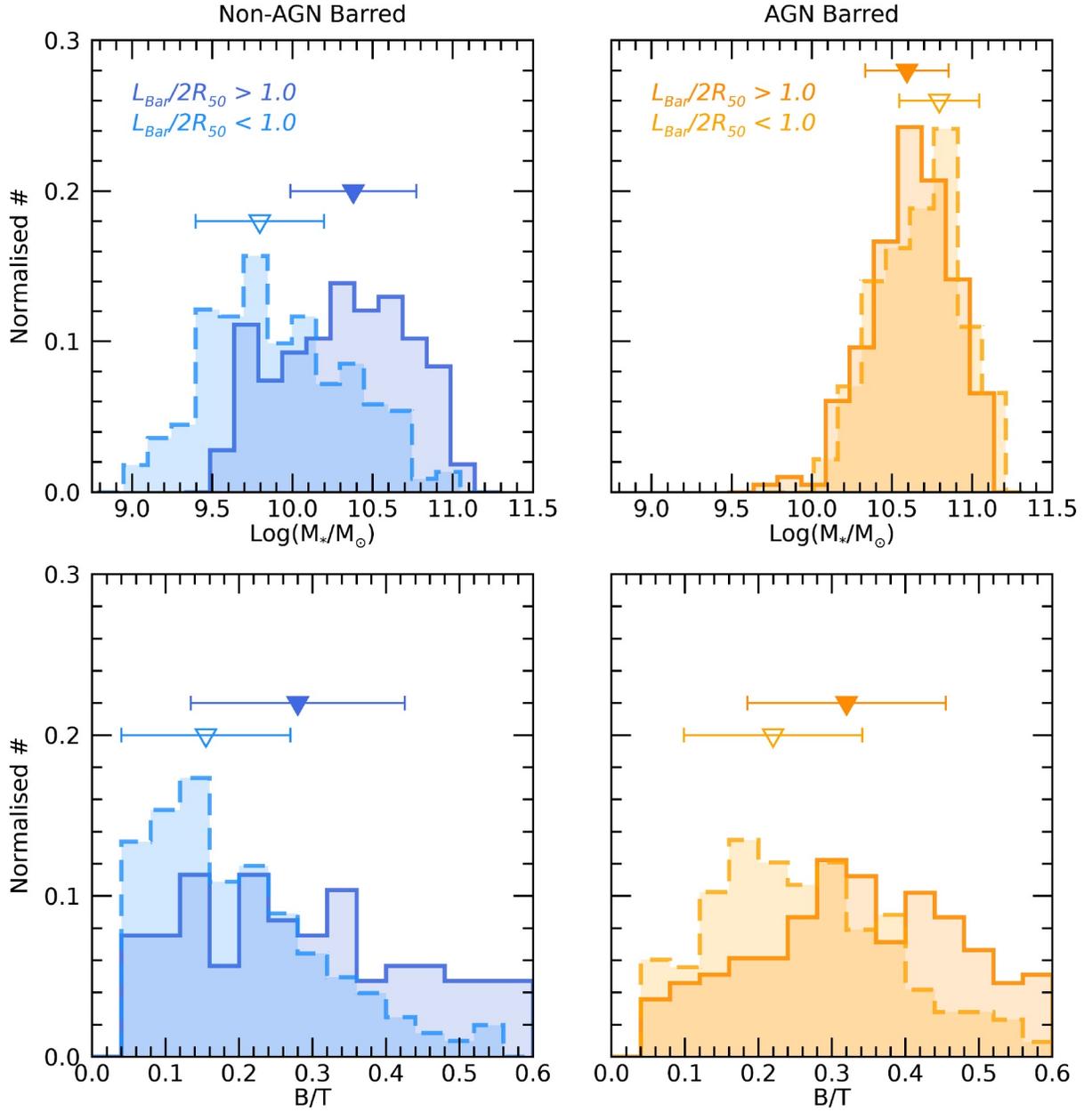

**Figure 7.** Upper panels: The histogram of stellar mass for non-AGN (left, blue colored) and AGN (right, orange colored) barred galaxies. Samples are divided into relatively longer ($L_{\rm Bar}/2R_{50} > 1.0$) and shorter ($L_{\rm Bar}/2R_{50} < 1.0$) bars. Darker-colored solid lines and lighter-colored dashed lines represent distributions of galaxies with longer and shorter bars, respectively. For comparison, the median value and standard deviation are shown as filled and open triangle symbols and error bars above each histogram. Lower panels: The same as upper panels but for the distribution of B/T.

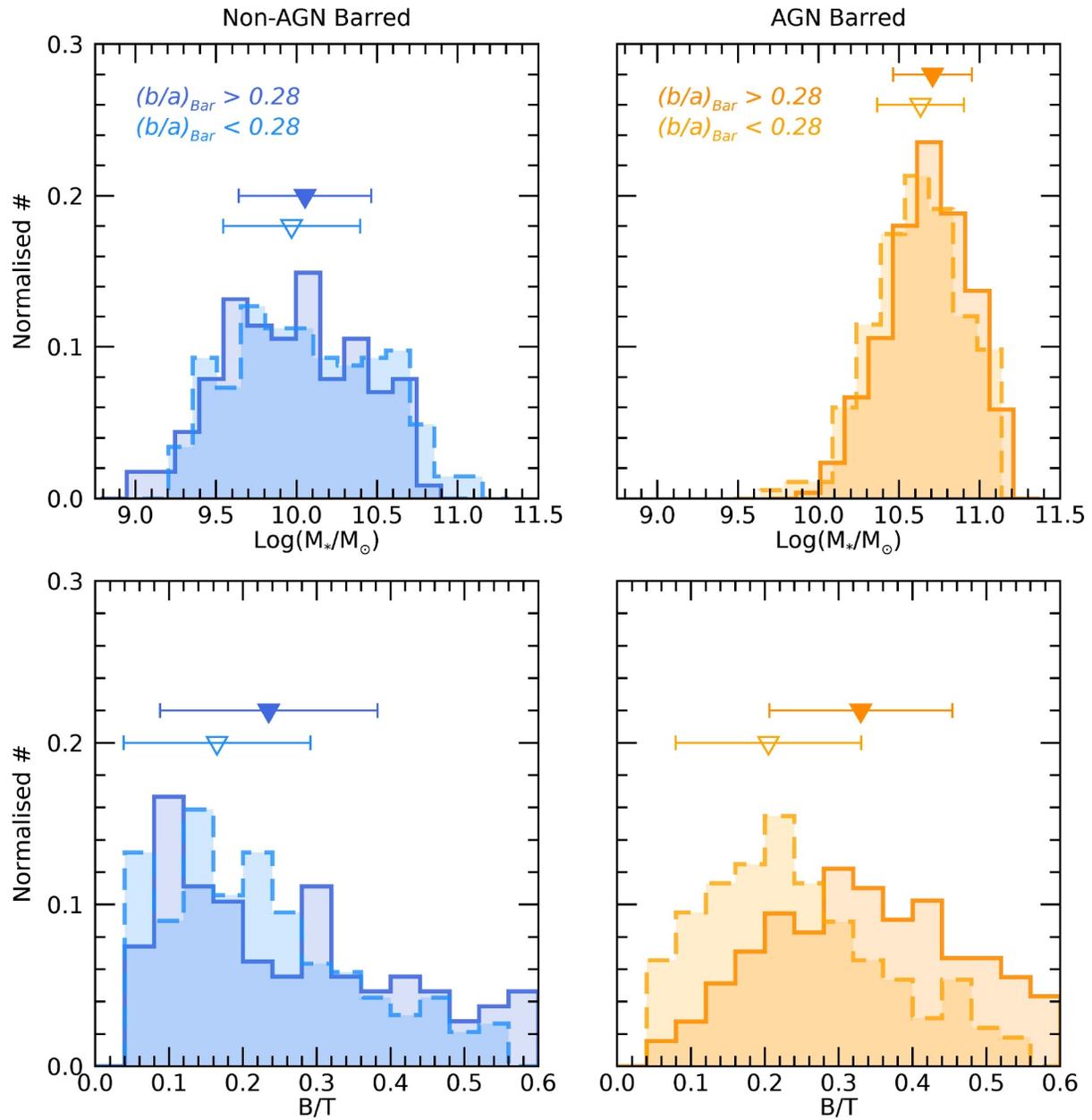

**Figure 8.** The same as Figure 7 but for the separation by bar's axis ratio, $(b/a)_{\rm Bar}$. Darker-colored solid lines and lighter-colored dashed lines represent the distribution of galaxies with rounder $((b/a)_{\rm Bar} > 0.28)$ and more elongated $((b/a)_{\rm Bar} < 0.28)$ bars, respectively.

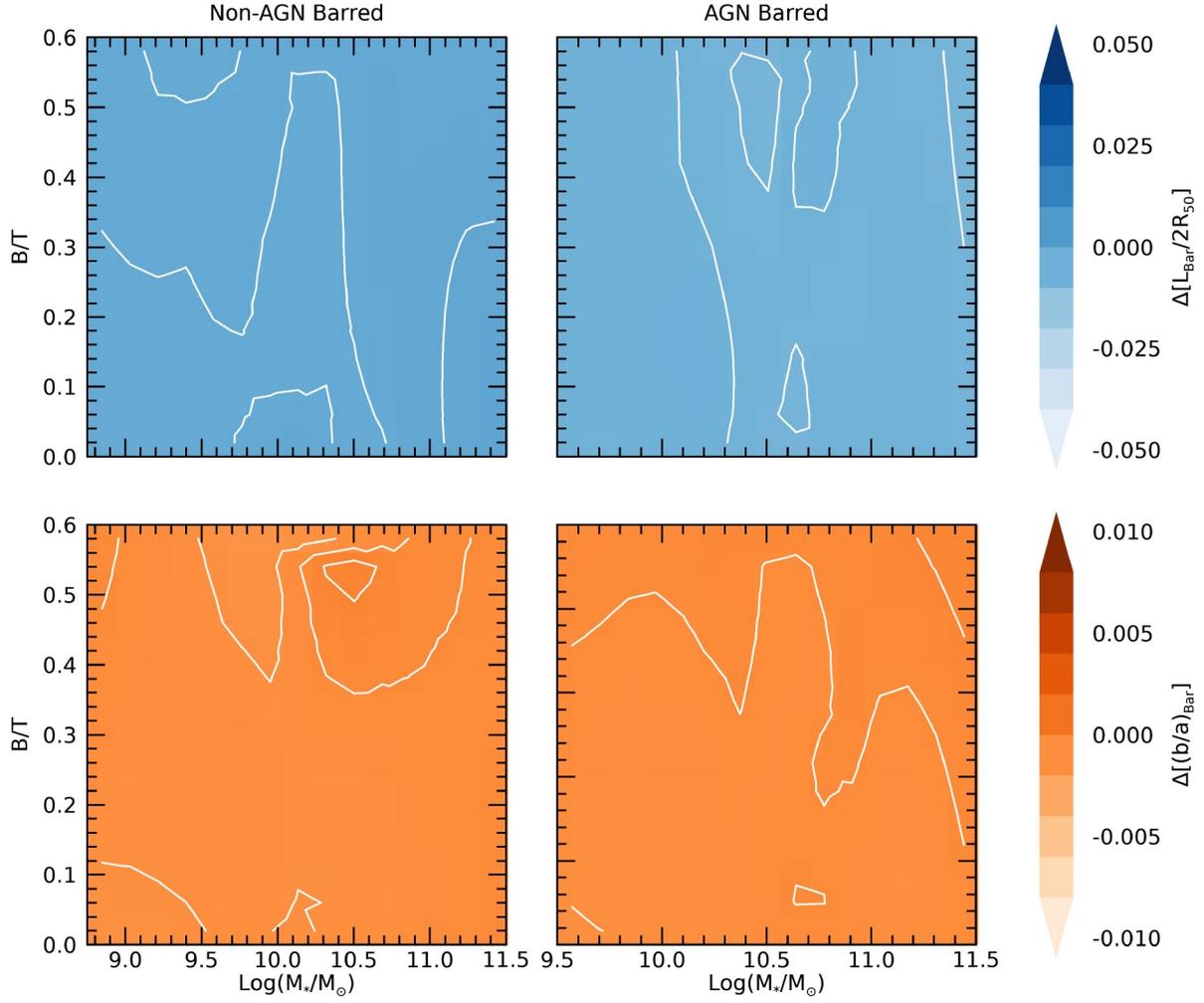

**Figure 9.** The same as Figure 6 but for the newly defined bar parameters, $\Delta\,[L_{\rm Bar}/2R_{50}]$ and $\Delta\,[(b/a)_{\rm Bar}]$, with more restricted ($\times\,0.1$) range of parameter values ([$-0.05$, $0.05$] for $\Delta\,[L_{\rm Bar}/2R_{50}]$ and [$-0.01$, $0.01$] for $\Delta\,[(b/a)_{\rm Bar}]$, respectively). The local regression scheme of Cappellari et al. (2013) is applied.

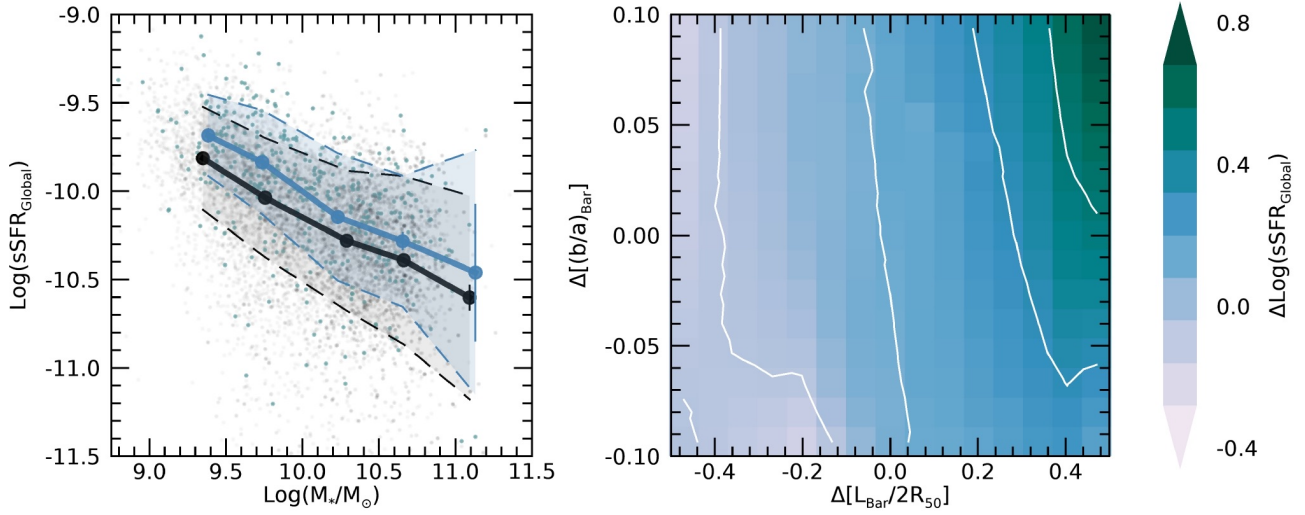

**Figure 10.** Left panel: The distribution of sSFR$_{\rm Global}$ of non-AGN galaxies as a function of stellar mass for unbarred (black) and barred (blue) ones. Colored dashed bands denote the standard deviation. The length of the error bars is the standard error, which is estimated as the standard deviation divided by the square root of the sample size at each stellar mass bin. Right panel: The residuals of sSFR$_{\rm Global}$, $\Delta{\rm Log(sSFR_{Global})}$, of non-AGN barred galaxies with respect to its corresponding unbarred control sample on the $\Delta\,[L_{\rm Bar}/2R_{50}]$ versus $\Delta\,[(b/a)_{\rm Bar}]$ plane. The local regression scheme of Cappellari et al. (2013) is applied.

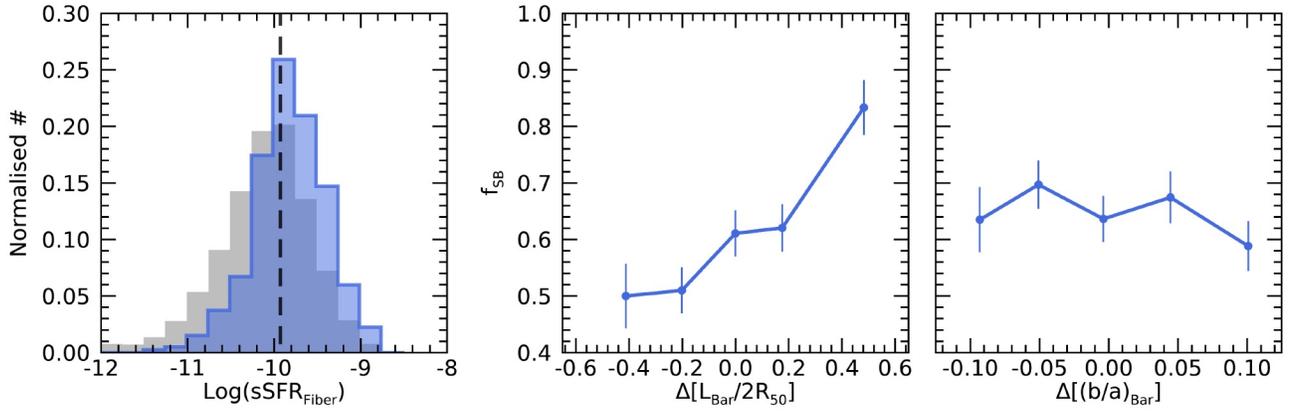

**Figure 11.** Left panel: The histogram of sSFR$_{\rm Fiber}$ of non-AGN galaxies for barred (blue) and unbarred ones (grey). The black dashed line represents the peak of the Gaussian fitted distribution of barred galaxies. Middle panel: The fraction of central starburst galaxies, $f_{\rm SB}$, as a function of $\Delta\,[L_{\rm Bar}/2R_{50}]$. The length of the error bars in each parameter bin represents the Poisson error for that bin. Right panel: The same as the middle panel but for the correlation between $f_{\rm SB}$ and $\Delta\,[(b/a)_{\rm Bar}]$

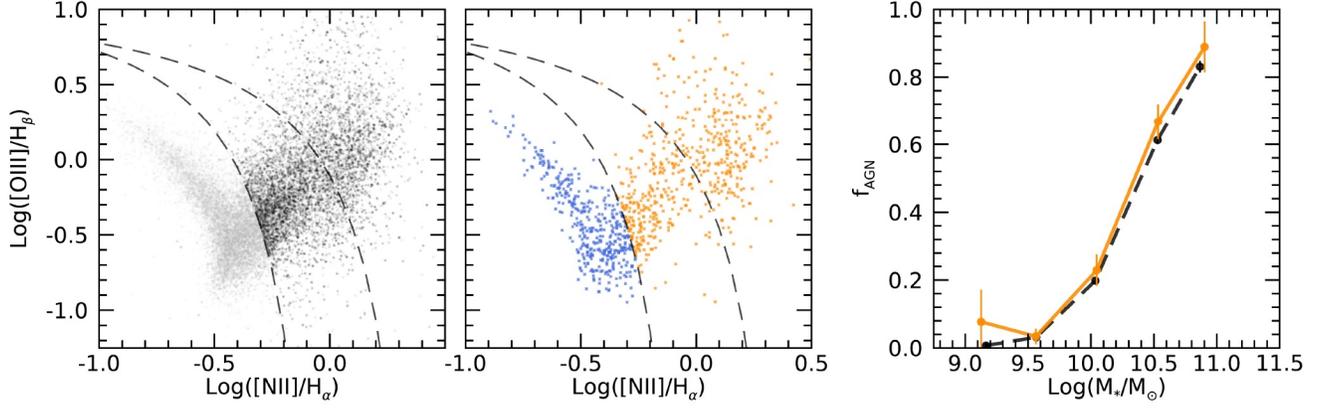

**Figure 12.** Left panel: The distribution of unbarred galaxies on the BPT diagram. The grey and black dots are non-AGN and AGN galaxies, respectively. Middle panel: The distribution of barred galaxies on the BPT diagram. The blue and orange dots are non-AGN and AGN galaxies, respectively. Right panel: The AGN fraction, $f_{\rm AGN}$, as a function of stellar mass for unbarred (black) and barred (orange) galaxies. The length of the error bars in each mass bin represents the Poisson error for that bin.

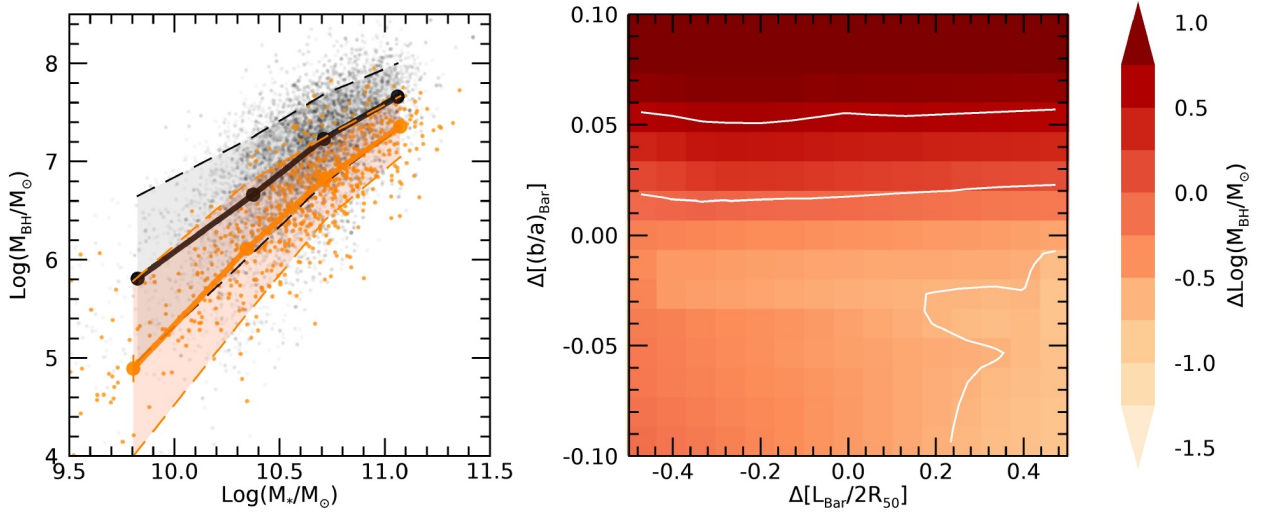

**Figure 13.** Left panel: The distribution of $M_{\rm BH}$ of AGN galaxies as a function of stellar mass for unbarred (black) and barred (orange) ones. Colored dashed bands denote the standard deviation. The length of the error bars is the standard error, which is estimated as the standard deviation divided by the square root of the sample size at each stellar mass bin. Right panel: The residuals of $M_{\rm BH}$, $\Delta {\rm Log}(M_{\rm BH}/M_*)$, of AGN barred galaxies with respect to its corresponding unbarred control sample on the $\Delta\,[L_{\rm Bar}/2R_{50}]$ versus $\Delta\,[(b/a)_{\rm Bar}]$ plane. The local regression scheme of Cappellari et al. (2013) is applied.

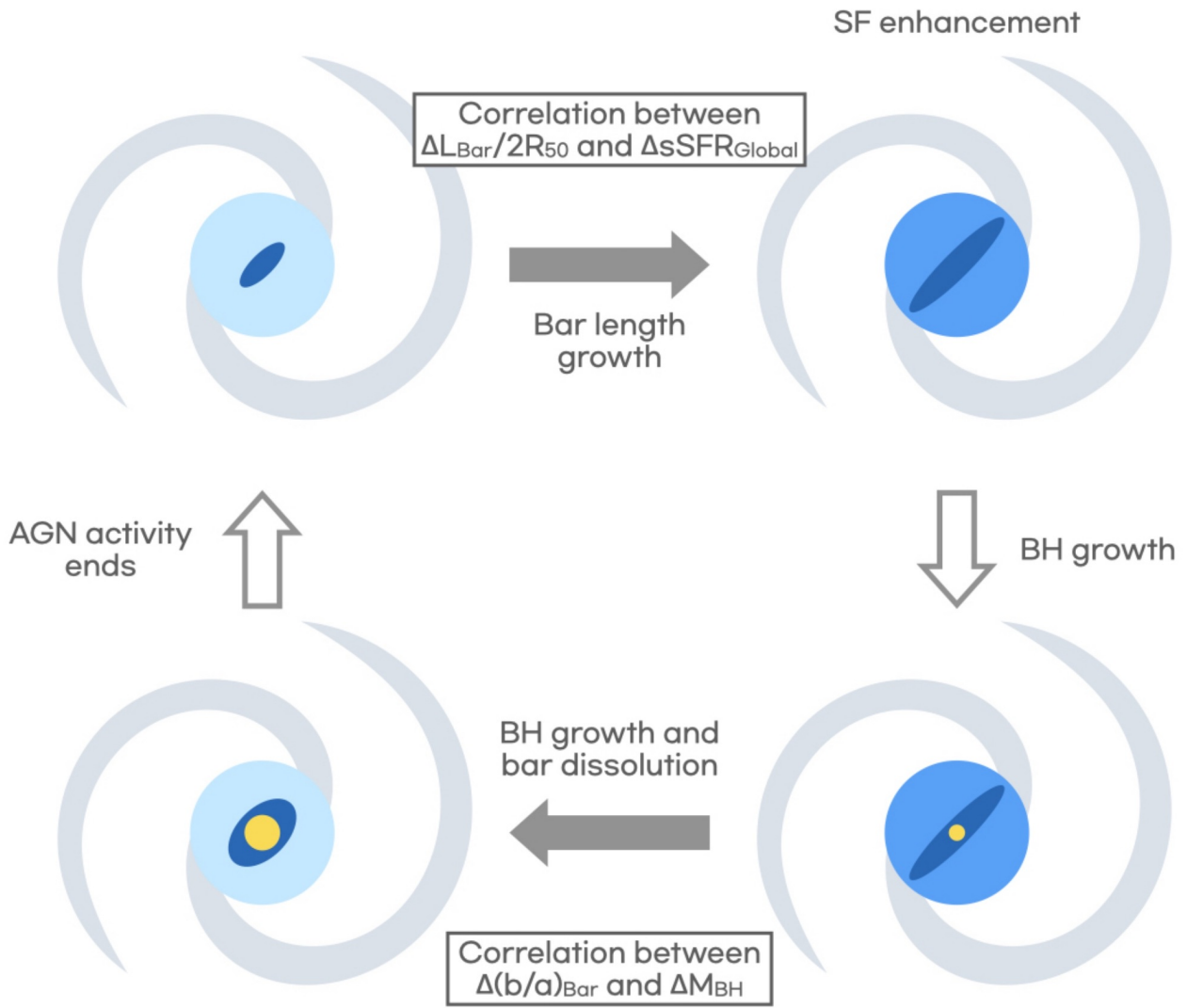

**Figure 14.** A schematic diagram of the evolution of barred galaxies. Initially, bars (represented by the blue ellipse at the center) begin to elongate and increase their SF activities (indicated by the darker blue color). However, once BHs are formed at the center (represented by the yellow circle), the bars are destroyed and become more rounded as the BHs evolve. The growth of BH mass in barred galaxies is limited, resulting in the presence of less massive BHs in these galaxies compared to unbarred galaxies. The filled arrows indicate the theoretically expected chronological evolution of barred galaxies, which are consistent with the results of our study.